\documentclass[12pt]{iopart}
\pdfoutput=1
\usepackage{iopams,diagbox,stackrel}  
\usepackage[utf8]{inputenc}
\usepackage{amssymb}
\expandafter\let\csname equation*\endcsname\relax
\expandafter\let\csname endequation*\endcsname\relax

\usepackage{bbold}
\usepackage{amsmath}

\usepackage{slashed}
\usepackage{nicefrac}
\usepackage{hhline}
\usepackage{braket}

\usepackage{hyperref}
\usepackage[nosort]{cite}

\usepackage{amsfonts}
\usepackage{mathrsfs}

\usepackage{graphicx}
\usepackage{cite} 
\usepackage{cleveref}
\usepackage[usenames,dvipsnames,svgnames,table]{xcolor}
\usepackage[colorinlistoftodos]{todonotes}

\usepackage{etoolbox}

\makeatletter
\def\@mkboth#1#2{}
\newlength\appendixwidth
\preto\appendix{\addtocontents{toc}{\protect\patchl@section}}
\newcommand{\patchl@section}{%
  \settowidth{\appendixwidth}{\textbf{Appendix }}%
  \addtolength{\appendixwidth}{1.5em}%
  \patchcmd{\l@section}{1.5em}{\appendixwidth}{}{\ddt}%
}
\makeatother

\numberwithin{equation}{section}

\makeatletter
\long\def\@makefntext#1{\parindent 1em\noindent 
 \makebox[1em][l]{\footnotesize\rm$\m@th{^{\arabic{footnote}}}$}%
 \footnotesize\rm #1}
\def\@makefnmark{\hbox{$^{\arabic{footnote}}\m@th$}}
\def\@thefnmark{$^{\arabic{footnote}}$}
\makeatother

\newcommand\Tstrut{\rule{0pt}{2.8ex}}         

\newcommand{\A}{\mathbb{A}}
\newcommand{\C}{\mathbb{C}}

\newcommand{\CP}{\mathbb{CP}}

\newcommand{\PT}{\mathbb{PT}}

\renewcommand{\P}{\mathbb{P}}

\newcommand{\cH}{\mathcal{H}}
\newcommand{\scri}{\mathscr{I}}
\newcommand{\M}{\mathbb{M}}

\newcommand{\T}{\mathbb{T}}

\newcommand{\p}{\partial}
\newcommand{\dbar}{\bar\partial}

\newcommand{\cA}{\mathcal{A}}
\newcommand{\cB}{\mathcal{B}}

\newcommand{\cE}{\mathcal{E}}
\newcommand{\cI}{\mathcal{I}}
\newcommand{\cM}{\mathcal{M}}
\newcommand{\cN}{\mathcal{N}}
\newcommand{\cO}{\mathcal{O}}

\newcommand{\cV}{\mathcal{V}}

\renewcommand{\P}{\mathbb{P}}
\newcommand{\SL}{\mathrm{SL}}
\newcommand{\Pf}{\mathrm{Pf}}
\newcommand{\Pt}{\mathrm{PT}}

\newcommand{\GL}{\mathrm{GL}}

\newcommand{\rd}{\, \mathrm{d}}

\newcommand{\pf}{\mathrm{Pf}\,}
\newcommand{\be}{\begin{equation}}
\newcommand{\ee}{\end{equation}}
\newcommand{\bea}{\begin{eqnarray}\label}
\newcommand{\eea}{\end{eqnarray}}

\newcommand{\vol}{\mathrm{vol}\,}
\newcommand{\la}{\langle}
\newcommand{\ra}{\rangle}

\newcommand{\CHY}{{\scalebox{0.6}{CHY}}}
\newcommand{\sA}{{\scalebox{0.6}{$A$}}}

\newcommand{\sL}{{\scalebox{0.6}{$L$}}}
\newcommand{\sR}{{\scalebox{0.6}{$R$}}}

\crefformat{footnote}{#2\footnotemark[#1]#3}

\begin{document}

\begin{flushright}
	SAGEX-22-07
\end{flushright}

\title[Ambitwistor strings]{The SAGEX Review on Scattering Amplitudes \\ \vspace{10pt}
  Chapter 6: 
  Ambitwistor Strings and Amplitudes from the Worldsheet 
  }

\author{Yvonne Geyer}

\address{Department of Physics, Faculty of Science, Chulalongkorn University\\
Thanon Phayathai, Pathumwan, Bangkok 10330, Thailand}
\ead{yjgeyer@gmail.com}

\author{Lionel Mason}
\address{The Mathematical Institute, University of Oxford, \\
24-29 St.~Giles, Oxford OX1 3LP, United Kingdom}
\ead{lmason@maths.ox.ac.uk}
\vspace{20pt}

\begin{abstract}
Starting with Witten's twistor string, chiral string theories  have emerged that  describe   \emph{field theory} amplitudes without the towers of massive states of conventional strings. These models are known as ambitwistor strings due to their
target space; the space of  complexified null geodesics, also called ambitwistor space.
Correlators in these string theories directly  yield compact formul{\ae} for tree-level amplitudes  and  loop integrands, in the form of  worldsheet integrals fully localized on solutions to constraints known as the \emph{scattering equations}. 
In this chapter, we discuss two incarnations of the ambitwistor string:  a `vector representation' starting in space-time and structurally resembling the RNS  superstring, and a four-dimensional twistorial version closely related to, but distinct from Witten's original model.
The RNS-like models exist for several theories, with   `heterotic' and type II models describing super-Yang-Mills and 10d supergravities  respectively,
and they manifest the double copy relations directly at the level of the worldsheet models.
In the second half of the chapter, we explain how  the underlying models lead to diverse applications, ranging from extensions to new sectors of theories,  loop amplitudes and to scattering on curved backgrounds.  We conclude with a brief discussion of connections to conventional strings and celestial holography.
\end{abstract}

\newpage

\tableofcontents

%
%
%
%
%

\newpage


\section{Introduction}
The twistor-string of Witten and   Berkovits \cite{Witten:2003nn,Berkovits:2004hg,Berkovits:2004jj} 
marked a turning point in the study of scattering amplitudes, exposing systematic structures that are not apparent from either standard space-time Lagrangians or from conventional strings.   In conventional string theory, the target space is space-time, whereas for the twistor-string, it is the cotangent bundle of  twistor space, where twistor space $\PT$ is a three-complex-dimensional  manifold.  The string path integral localizes onto  holomorphic maps from a Riemann surface $\Sigma$, the string worldsheet, into $\PT$.  The theory led to formul{\ae} for the complete tree-level S-matrix for four-dimensional super Yang-Mills of unprecedented simplicity. These have by now been generalized to a great variety of theories and to loop integrands and beyond.  This review will cover many of these developments.  By way of introduction we give a brief history.

\subsection{A brief history}

The story starts with Nair's reformulation \cite{Nair:1988bq} of the famous Parke-Taylor formula for the colour-ordered MHV Yang-Mills amplitude
\begin{equation} \label{Parke-Taylor}
A^{\rm MHV}(1^+, \dots, i^-, \dots, j^-, \dots, n^+)=\frac{\langle{ij\rangle}^4}{\langle 12\rangle\langle 23\rangle \dots \langle(n-1) n\rangle \langle n1\rangle}\, ,
\end{equation}
where the $n$ particles have massless momenta $k_i=|i]\langle i|$ in spinor helicity notation, and particles $i,j$ have negative helicity with the rest positive.
Nair, in an elegant $\cN=4$ supersymmetric formulation including the supermomentum conserving delta-function, expressed this
as an integral of a current-algebra correlator   over the  moduli space of Riemann spheres, $\CP^1$s, holomorphically  embedded in supertwistor space of degree one (i.e., lines in $\CP^3$).  In the twistor string \cite{Witten:2003nn,Berkovits:2004hg,Berkovits:2004jj}, 
  N$^{k-2}$MHV amplitudes with  $k$ negative helicity gluons arise as  integrals over the moduli space of degree $k-1$ curves in $\PT$.

 A striking output was  the  formula of Roiban, Spradlin and Volovich (RSV) \cite{Roiban:2004yf}.   They were able to perform some of the moduli integrals 
so as to  express the full tree-level S-matrix for $\cN=4$ super Yang-Mills tree-amplitudes 
as a sum over  residues.   It was soon observed  \cite{Witten:2004cp} that the equations that determine these residues are underpinned by  the \emph{scattering equations}.  These equations determine  $(n-3)!$ sets of $n$ points $\{\sigma_i\}\subset \CP^1$ up to M"{o}bius transformations, i.e. points in  the moduli space  $\mathfrak{M}_{0,n}$, in terms of the $n$ null momenta $k_i$ of the scattering particles: 
\begin{equation}\label{SE}
\cE_i:= \sum_{j=1}^n \frac{k_i\cdot k_j}{\sigma_i-\sigma_j}=0\, .
\end{equation}
These equations play a pivotal role in the subsequent story.

It emerged that the  twistor-string is equivalent  to $\cN=4$ super Yang-Mills coupled to a certain conformal supergravity \cite{Berkovits:2004jj}.  This does imply  that attempts to compute Yang-Mills loop amplitudes via this method would be corrupted by conformal supergravity modes running in the loops; these  are not in any case easy to compute \cite{Dolan:2007vv}.

There were two lines of attack to find analogous formul{\ae} and theories for Einstein gravity, one by improving our understanding of its  MHV amplitude and the other via the double copy \cite{Bern:2010ue,Chapter2}. The latter expresses Einstein gravity amplitudes as a `square' of the different colour-ordered Yang-Mills amplitudes using an inner product, the `KLT' momentum kernel, introduced by Kawai, Lewellen and Tye \cite{Kawai:1985xq}.  
Following  \cite{Cachazo:2012uq} it was conjectured  that  the residues on which the RSV formul{\ae} are supported in fact diagonalize the KLT kernel,  so that the double copy can be implemented  on the RSV formula to produce amplitudes for Einstein gravity in four dimensions \cite{Cachazo:2012da}.  On the other hand, Hodges had found an optimal version of the MHV formula for gravity amplitudes \cite{Hodges:2012ym} in terms of reduced determinants that bore little relation to these formul{\ae}. This led Cachazo and Skinner to introduce a compact worldsheet formula for $\cN=8$ supergravity amplitudes  \cite{Cachazo:2012kg,Cachazo:2012pz} that was soon seen to arise from a twistor-string for $\cN=8$  supergravity \cite{Skinner:2013xp}.  In this theory,   Hodges' reduced determinants and their generalizations are obtained from fermion correlators on the worldsheet.

Cachazo, He and Yuan (CHY) pursued the   relationship between the double copy and the scattering equations,  proving that the solutions to the scattering equations do indeed diagonalize the KLT kernel and giving an elegant formulation for its expression as a reduced determinant on $\mathfrak{M}_{0,n}$, \cite{Cachazo:2013gna}.   Both the double copy and the scattering equations make sense  in arbitrary dimension and the CHY collaboration were soon able to produce expressions for Yang-Mills and gravity amplitudes in all dimensions \cite{Cachazo:2013hca} that perfectly express the double copy within formul{\ae} of the form
\begin{equation}\label{CHY}
\cA=\int \cI^\sL\cI^\sR\, \frac{\prod_{i=1}^n\bar\delta(\cE_i)\, d\sigma_i}{\mathrm{vol \, PSL}(2,\C)\times \C^3}\, .
\end{equation}
Here the $\bar\delta(\cE_i)$ are delta-functions that impose the scattering equations (\ref{SE}) and the  PSL$(2)$ quotient by M\"{o}bius transformations on the $\sigma_i$  is intended in the Faddeev-Popov sense, reducing the formula to an $(n-3)$-dimensional integral  over $\mathfrak{M}_{0,n}$, the moduli space of $n$ marked points on the Riemann sphere.  This then localizes on $(n-3)!$ residues at the solutions to the scattering equations (\ref{SE}). 

The factors $\cI^{\sL/\sR}$ are theory dependent, and can depend on polarization or colour data associated to particles in the theory; this factorization then gives diagonalized 
expression of the double copy.  The zero'th copy is when both are associated to colours.  For two different colour orderings, they are  given by Parke-Taylor  expressions each with denominators like that of (\ref{Parke-Taylor}) but with $\langle ij\rangle$ replaced by $\sigma_i-\sigma_j$.  The amplitudes are then  those of  a theory of biadjoint scalars
 \cite{Cachazo:2013iea}, i.e., $\phi^{\mathfrak{a}\tilde{\mathfrak a}}(x)$ with $\mathfrak a$ being an index associate to a Lie algebra $\mathfrak{g}$ and $\tilde{\mathfrak a}$ associated to another $\tilde{\mathfrak{g}}$; 
 the theory has cubic interactions  $f_{\mathfrak{abc}}\tilde f_{\tilde{\mathfrak a} \tilde{\mathfrak b} \tilde{\mathfrak c}} \, \phi^{\mathfrak{a}\tilde{\mathfrak a}}\phi^{\mathfrak{b}\tilde{\mathfrak b}} \phi^{\mathfrak{c}\tilde{\mathfrak c}}$ determined by the two sets of structure constants $f_{\mathfrak{abc}}$ and $\tilde f_{\tilde{\mathfrak a} \tilde{\mathfrak b} \tilde{\mathfrak c}}$.   If one Parke-Taylor is replaced by a certain reduced Pfaffian, see \eqref{Pfaffian}, then the formula yields Yang-Mills amplitudes, and if both are such Pfaffians, one obtains gravity formul{\ae}.  CHY soon extended their framework to include remarkable new expressions for amplitudes of many more theories of interest such as effective theories, nonlinear-Sigma models,  Born-Infeld and so on  \cite{Cachazo:2014nsa, Cachazo:2014xea}.

These formul{\ae} have undoubted theoretical importance in their own right, but the question remained as to what physical principles generate them; they look unlike anything that arises from a space-time action formulation and the conventional string does not localize on residues in the field theory limit. 
Such underlying principles should  for example give insights into extensions to loop amplitudes or non-pertubative phenomena.  
 Both the CHY formul{\ae} and  twistor-strings are now  understood under the umbrella of  ambitwistor-string theories \cite{Mason:2013sva}.  
These  are quantum field theories of holomorphic maps from a Riemann surface to ambitwistor space, $\A$; this is defined to be the complexification of the phase space of a massless particle.  In four dimensions, $\A$ can be realized as the cotangent bundle of twistor space, $\A=T^*\PT$ and of its dual $\A=T^*\PT^*$; twistor space is chiral, and dual twistor space antichiral, so $\A$ is ambidextrous, hence its name.\footnote{ It was introduced by  Witten and Isenberg,as space on which one can encode general (super) Yang-Mills fields \cite{Witten:1978xx, Isenberg:1978kk} generalizing Ward's twistor  construction \cite{Ward:1977ta} for self-dual Yang-Mills fields.}  The original twistor-strings can be of either chirality but can both be understood in this way,  and indeed there is an ambidextrous version  \cite{Geyer:2014fka} in the same twistor coordinates that generates  formul{\ae} that are distinct from those of RSV and Cachazo-Skinner and we introduce these in \S\ref{sec:twistor-models}.
However, the simplest presentation that connects most directly to the CHY formul{\ae} is a presentation of ambitwistor space analogous to the original Ramond Neveu-Schwarz (RNS)  model for the conventional string and so we start with this in the next section \S\ref{sec:models}.  In particular in \S\ref{sec:proof} we give a more complete proof of the CHY formul{\ae} by BCFW recursion than is easy to find in the literature.

The second half of the review focuses on one of the key applications of the ambitwistor-string framework, the extension of the tree-level formul{\ae} to those that provide loop integrands. In \S\ref{Loops} we see that following the usual string paradigm, loop integrands can be constructed via higher genus worldsheets.  However, this yields  formul{\ae} that are at least superficially highly transcendental for loop integrands that should be rational functions. In \S\ref{nodal} we explain how, by means of a residue theorem, such formul{\ae} can be reduced to ones based on nodal Riemann spheres.  We go on to explain various new representations of loop integrands at one and two loops and further applications to the double copy.  In the final \cref{Frontiers} we briefly discuss further frontiers,   extensions to curved backgrounds, and connections  with the conventional string and with celestial holography, providing pointers to the literature.


\section{Ambitwistor geometry and models}
\label{sec:models}

Ambitwistor-string theories are chiral strings, i.e., quantum theories of holomorphic maps  from a Riemann surface $\Sigma$ into  a complex manifold.  The target space,  ambitwistor space $\A$, is   the complexification  of the real phase space  of null geodesics.  In $d$ dimensions, points of $\A$ correspond to  complex null geodesics in a  complexified space-time $(M,g)$ in which the metric depends holomorphically on the $d$ complex coordinates on $M$: this can be obtained by complexification of a real space-time with analytic metric.

There are many real worldline actions for massless particles with different couplings to background fields and supersymmetry  and there is a simple recipe to go from such a real worldline action to a complex  ambitwistor-string action.  Here we start with the simplest  first--order version  in a $d$-dimensional space-time $(M,g)$:
\be\label{null-particle}
S[X,P]=\frac{1}{2\pi}\int P_\mu\rd X^\mu -  \frac{\tilde e}{2} g^{\mu\nu}P_\mu P_\nu\, .
\ee
In this action, the einbein $\tilde e$ is a Lagrange multiplier enforcing the constraint $P^2=0$, and is also the worldline gauge field for the gauge transformations\footnote{for simplicity these are given for flat space; in curved space we must include Christoffel symbols.}
\be
	\delta X^\mu = \alpha \,g^{\mu\nu}P_\nu\qquad \delta P_\mu=0 \qquad \delta e=\rd\alpha \,,
\ee
conjugate to this constraint. Thus  $P$ must be null and gauge transformations send fields $X$ to  $X'$ along the null translation generated by $P$. The solutions to the field equations modulo   gauge are null geodesics in space-time, parametrized by the scaling of $P$. The quantization of this action leads to the massless Klein-Gordon equation.

The ambitwistor string replaces  the worldline with a Riemann surface $\Sigma$ and  complexifies  the target space so that the $(P_\mu, X^\mu)$ are holomorphic coordinates on the cotangent bundle $T^*M$ of a complexified space-time $(M,g)$.  The algorithm to obtain an ambitwistor-string starts by replacing $\rd X$ in~\eqref{null-particle} by 
\begin{equation}
\bar \p_{e}  X = \rd\bar\sigma \, \p _{\bar\sigma} X - e \, \p_\sigma X\,,
\end{equation}
 to obtain the bosonic action
\be\label{boson-str}
S_{\mathrm{bos}}[X,P]=\frac1{2\pi}\int_\Sigma  P_\mu \bar \p_{e}  X^\mu -  \frac{\tilde e}{2} g^{\mu\nu}P_\mu P_\nu\, .
\ee
For the kinetic term of~\eqref{boson-str} to be invariant, we must take $P_\mu$ to be a complex (1,0)-form, i.e.,  with values in the \emph{canonical bundle} $K_\Sigma:=\Omega^{1,0}_\Sigma$ on the worldsheet.  Thus,   when $ e=0$, $P_\mu = P_{ \sigma\,\mu}\rd\sigma$ where $\sigma $ is  a local holomorphic worldsheet coordinate. We must take both $e$ and $\tilde e$  to be  $(0,1)$-forms on $\Sigma$ with values in holomorphic vector fields $T_\Sigma$, i.e.,  Beltrami differentials.
The worldsheet field $e$ plays the same role as in the conventional string,  parametrizing complex structures on $\Sigma$ and gauging worldsheet diffeomorphisms, but now in a chiral model. On the other hand, the geometric interpretation of $\tilde e$ is quite different from that of the ordinary string.  It imposes the constraint $P^2=0$ 
and  gauges the transformations
\be\label{gauge}
	\delta  X^\mu = \alpha \,g^{\mu\nu}P_\mu\qquad \delta P_\mu=0 \qquad \delta \tilde e = \bar \p  \alpha\,.
\ee
 Here we must take  $\alpha$ to transform as a holomorphic vector on the worldsheet. Thus 
 $(X,P)$ describe  a map into complexified cotangent bundle $T^*M$ of complexified space-time, but imposing the constraint $P^2=0$ and quotienting by the gauge symmetry generated by the geodesic spray $P\cdot \p_X$ reduces the target space of~\eqref{boson-str} to the space of complex null geodesics $\A$, ambitwistor space, via

\begin{equation}
p: T^*M\big|_{P^2=0} \;\;\; \longrightarrow \;\;\; \A:=\,T^*M\big|_{P^2=0}\;\big/\; P\cdot\p_X\, .
\end{equation}
Unlike the particle case, $P_\sigma$ is only defined up to a rescaling ($P$ takes values in the canonical bundle of $K_\Sigma$) so there is no preferred scaling of these geodesics, reducing the target space further to projective ambitwistor space $\P\A$.

Following the double copy, all our models will take the form 
\begin{equation}
S=S_{\mathrm{bos}}+ S^\sL +S^\sR\, ,
\end{equation}
where $S^L$ and $S^R$ are two independent choices of worldsheet matter. There are a number of interesting choices outlined in \cite{Casali:2015vta}, but  in order to establish the basic models, we will focus on the following two worldsheet systems:

\paragraph{Current algebras.} The first will be a current algebra with action  denoted by $S_C$.  It provides  `currents' $j^a \in K_\Sigma \otimes \mathfrak{g}$, where $a$ is a Lie algebra index associated to the Lie algebra $\mathfrak{g}$, that  satisfy the OPE
 \begin{equation}
 j^a(\sigma) j^b(\sigma')\sim  \frac{k\,\delta^{ab}}{(\sigma-\sigma')^2}+ \frac{f^{ab}_c j^c}{\sigma-\sigma'}\, .
 \end{equation}
Here $\delta^{ab}$ is the Killing form and $  f^{ab}_c$ the structure coefficients of $\mathfrak{g}$, and $k$ the level of the current algebra.  The simplest $S_C$  arises from  free fermions $\rho^{\alpha}\in K^{\nicefrac{1}{2}}_\Sigma\otimes \C^N$ with action
\begin{equation}
S_\rho :=\int \rho^{\alpha}\bar\p \rho^\beta\,\delta_{\alpha\beta}\, , \qquad \leadsto \qquad 
j^{\alpha\beta}:= \rho^{\alpha}\rho^\beta \in K\otimes \mathfrak{so}(N)\, , \quad k=1\,. 
\end{equation}
We will not in general specify the action in detail, and merely denote it by $S_C$.

\paragraph{Worldsheet superalgebra.} This system consists of fermionic spinor $\Psi^\mu\in K^{\nicefrac{1}{2}}\otimes \C^d$ as a fermionic counterpart to $X^\mu$, and a fermionic gauge field $\chi\in \Omega^{0,1}\otimes T^{\nicefrac{1}{2}}_\Sigma$ with action   
\begin{equation}
S_\Psi= \int g_{\mu\nu}\Psi^\mu \dbar \Psi^\nu - \chi P\cdot \Psi\, .
\end{equation}
 The field $\chi$ is a gauge field generating degenerate fermionic  supersymmetries
 \begin{equation}\label{eq:sym_chi}
 \delta X^\mu = \epsilon \Psi^\mu
 \qquad 
 \delta \Psi^\mu = \epsilon P^\mu
 \qquad
 \delta P_\mu = 0
 \qquad
 \delta \chi= \dbar \epsilon\, .
\end{equation}  

 With these ingredients there are three main consistent models mirroring the closed string models in conventional string theory:
\begin{itemize}
\item 
The biadjoint scalar model  is the bosonic model  with two current algebras
\begin{equation}
S_{\mathrm{BAS}}:=S_{\mathrm{bos}}+S_C +S_{\tilde C}\,,
\end{equation}
The two current algebras, $S_C+S_{\tilde C}$ provide currents $j^a$, $\tilde j^{\tilde a}$ respectively and generate bi-adjoint scalar amplitudes correctly. 
\item The heterotic model has one fermionic matter system $S_\Psi$ and one current algebra $S_C$, and generates Yang-Mills amplitudes correctly;
\begin{equation}
S_{\mathrm{het}}:=S_{\mathrm{bos}}+ S_\Psi + S_C\,.
\end{equation}
\item The type II models with two fermionic worldsheet matter systems $S_\Psi$, $S_{\tilde \Psi}$ generate supergravity amplitudes
\begin{equation}
S_{\mathrm{II}}:=S_{\mathrm{bos}}+ S_\Psi + S_{\tilde \Psi}\, .
\end{equation}
\end{itemize}
These models already manifest the double copy, and a naive\footnote{Naive because of the absence of Jacobi relations, or a suitable analogue of identical relations between $S_C$ and $S_\Psi$, see also the discussion in \cref{fdev}.} 
version of the colour-kinematics duality via the interchangeability of the current algebras with the $S_\Psi$ systems. In the double copy, the biadjoint scalar is the zeroth copy, Yang-Mills the single copy and gravity the double copy.

One can construct many further models with more elaborate choices of worldsheet matter and all models seem to give rise to amplitudes of some field theory, at least at tree level.  
These amplitude formul{\ae} include Born-Infeld, Dirac-Born-Infeld, Einstein-Yang-Mills, harmonic maps and more, manifesting a more extended double copy, see table \ref{models} and \cite{Cachazo:2014nsa, Cachazo:2014xea, Casali:2015vta}.
 Only the last models $S_{\mathrm{II}}$ correspond to conventional supergravity, yielding IIA or IIB supergravity in 10 dimensions according to the choice of GSO projection as in conventional string theory.
 
\subsection{BRST gauge fixing and quantization}
On the Riemann sphere in the absence of vertex operators, we can gauge fix by setting each of the gauge fields to zero (more generally 
we can fix the gauge fields to lie within the cohomology class $H^{0,1}(\Sigma,\,T_\Sigma(-\sigma_1-\ldots-\sigma_n)$).

After gauge fixing a ghost system is introduced for each gauge field, the fermionic $(b,c)\in K_\Sigma^2 \oplus T_\Sigma$ for $e$,  and $(\tilde b, \tilde c)\in K^2_\Sigma\oplus T_\Sigma$ for $\tilde e$, and the bosonic $(\beta, \gamma)\in K^{\nicefrac{3}{2}}_\Sigma\oplus T^{\nicefrac{1}{2}}_\Sigma$ for $\chi$ with free ghost actions
\begin{equation}
S_{(b,c)}=\int b\dbar c\, , \quad S_{(\tilde b,\tilde c)}=\int \tilde b\dbar \tilde c\, ,\quad S_{(\beta,\gamma)}=\int \beta\dbar \gamma\, .
\end{equation}
Invariance under the gauge symmetries is then imposed by considering the cohomology associated with the BRST operator $Q$ which takes the 
 form (here for the type II models)\footnote{For a general gauge system  generated by  currents $j^a$ of perhaps different spins or statistics, we have ghosts $c^a$ of opposite statistics and $Q=  \oint c_a j^a + \frac{1}{2} b^c f_c^{ab}c_ac_b$,  where $f^{ab}_c$ are  the structure constants.    }
\begin{equation}
Q=\oint  c\,  \Big(T^m+\frac{1}{2}\,T^{bc}\Big) +  \frac{\tilde c}{2} P^2 + \gamma\, P\cdot \Psi +\tilde \gamma\, P\cdot \tilde\Psi + \frac{1}{2}\, \tilde b \Big(\gamma^2 + \tilde\gamma^2\Big)\, ,
\end{equation}
where $T^m$ is the holomorphic part of the stress-energy tensor, and $T^{bc}=(\partial b) c+2b(\partial c)$.
Classically $Q^2$ vanishes by construction, but, as in standard string theory, the quantum models can be inconsistent as double contractions can give $Q^2\neq 0$.  We have

\begin{itemize}
\item
The pure bosonic model  $S_{\mathrm{bos}}$
 is critical in 26 dimensions.  However, for $S_{\mathrm{BAS}}=S_{\mathrm{bos}}+S_C +S_{\tilde C}$, the critical dimension will depend on the central charges of $S_C +S_{\tilde C}$.
\item The model $S_{\mathrm{het}}$
is critical in 10d with current algebras 
for $E_8\times E_8$ or SO$(32)$.

\item The type II models are critical in 10d.

\end{itemize}
The central charge calculations  are  analogous to those in conventional string theory.  Note that,  even when $Q^2\neq  0$, tree-level amplitude formul{\ae} generally make sense even though the underlying ambitwistor-string is not critical.

\subsection{Vertex operators}\label{sec:VO}
In string theory, amplitudes are constructed as correlation functions of vertex operators, with each vertex operator corresponding to an external particle. Vertex operators come in various \emph{pictures} that depend on how residual  gauge freedom is fixed after initial gauge fixing \cite{Green:1987a, Green:1987b,  Witten:2012bh}.  For worldsheet diffeomorphisms, the generic case for multiparticle amplitudes are \emph{integrated vertex operators} that require integration over $\Sigma$; these can be understood as the perturbations of the action corresponding to infinitesimal background plane-wave fields.  One also needs a small number of \emph{fixed vertex operators} that correspond to the same type of particles, but fix the residual gauge symmetries. These two types of vertex operators are related by pairing the fixed vertex operator with moduli insertions from the gauge-fixing procedure.\footnote{\label{fn:PCOs}From the CFT perspective, the fixed vertex operators are more fundamental, and correlators can be equivalently expressed using fixed vertex operators only, with additional moduli insertions often referred to as picture changing operators (especially for fermionic symmetries such as \eqref{eq:sym_chi}).}
We here give the basic recipes required for the amplitude formul{\ae} together with some brief intuitions on the geometry  following \cite{Mason:2013sva,Casali:2015vta}.  More sophisticated  derivations are  given in \cite{Adamo:2013tja, Ohmori:2015sha}.

\paragraph{Integrated vertex operators.} 
Space-time fields can be represented on ambitwistor space  via the \emph{Penrose transform}.  This realizes spin $s$  fields on space-time as cohomology classes $H^1(\P\A, \cO(s-1))$ on projective ambitwistor space $\P\A$: these  classes can be represented as $\dbar $-closed $(0,1)$-forms on $\P\A$  of homogeneity degree $s-1$ in $P$.  For spin $s$ plane-waves of the form $\epsilon_{\mu_1} \ldots \epsilon_{\mu_s}\e^{ik\cdot X}$, 
such cohomology classes can be written explicitly  as
\begin{equation}\label{drep}
(\epsilon\cdot P)^s \,\bar \delta (k\cdot P) \,\e^{ik\cdot x}\in H^1(\P\A,\cO(s-1))\, .
\end{equation}
Here, 
we define the complex delta function $\bar \delta (z)$ for a complex variable $z$ by
\begin{equation}
\bar \delta (z):=\dbar \, \frac{1}{ z}= 2\pi i \delta(\Re z) \delta(\Im z) d\bar z\, .
\end{equation}
Although expressed on $T^*M$,  the plane wave representative descends to $\P\A$ as  $k\cdot P=0$ on the support of the delta function, so that under $X\rightarrow X+ \alpha P$, $k\cdot X$ doesn't change.
For $s=1$ this provides the Maxwell version of the ambitwistor Yang-Mills correspondence of   Witten and  Isenberg, et.\ al.\ \cite{Witten:1978xx, Isenberg:1978kk} and for $s=2$ this provides the linear version of the transform for gravity introduced by Lebrun \cite{LeBrun:1983}, see \cite{ Baston:1987av,Mason:2013sva} for  general linear fields.

More generally, our integrated vertex operators all take the form
\begin{equation}
\cV:=\int w\, \bar{\delta}(k\cdot P)\,\e^{ik\cdot x}\, ,
\end{equation}
where $w$ depends on $P$ and the worldsheet matter fields from $S^\sL+S^\sR$.  With our identification of $\cO(1)=K_\Sigma$ on the worldsheet, $w \in K^2_\Sigma$, and the integrand defines a $(1,1)$-form as $\bar\delta(k\cdot P)$ has weight $-1$.
In general, to manifest the double copy,  we take 
\begin{equation}
w=v^\sL v^\sR\,, \qquad v^\sL, v^\sR \in K_\Sigma\, ,
\end{equation}
and $v^\sL$, $v^\sR$ are either of the form $t\cdot j$ where $j$ is a current associated to $S_C$ and $t^{\mathfrak{a}}\in \mathfrak{g}$, or $v^{\sL,\sR}=\epsilon\cdot P+  k\cdot \Psi\, \epsilon\cdot \Psi $ when associated with $S_\Psi$.  For the models that give the original CHY formul{\ae} for biadjoint scalar amplitudes, Yang-Mills and gravity, this yields\begin{align}
w_{\mathrm{BAS}}&:= t\cdot j\,\tilde t \cdot \tilde j, && \!\!t^{\mathfrak{a}}\in \mathfrak{g}\, , \quad \tilde t^{\dot{\mathfrak{a}}}\in \tilde{\mathfrak{g}}\, ,\nonumber \\
w_{\mathrm{YM}}&:= (\epsilon\cdot P + k\cdot \Psi \,\epsilon\cdot \Psi) \, t\cdot j\, , &&\label{int-V-ops}\\
w_{\mathrm{grav}}&:= (\epsilon\cdot P + k\cdot \Psi\, \epsilon\cdot \Psi )\,(\tilde\epsilon\cdot P +  k\cdot \tilde\Psi\, \tilde{\epsilon}\cdot\tilde \Psi)\, .&& \nonumber
\end{align}
The corresponding vertex operators are required to be $Q$-closed as part of the BRST quantization. Classically, this is automatic, but quantum mechanically, double contractions with the  $P^2$ term imply that $k^2=0$. In those models containing $S_\Psi$,  double contraction with the $P\cdot \Psi$ term further impose that $k\cdot \epsilon=0$.  Thus quantum consistency implies that our vertex operators correspond to on-shell fields in Lorentz gauge.  (The on-shell condition is \emph{not} a consequence of the classical ambitwistor Penrose transform.)  This gives the correct linear theory for bi-adjoint scalar and Yang-Mills theory.  
In the case of the (Q-invariant)  gravity vertex operators, the polarization vectors give the on-shell polarization data for a linearized metric, $g_{\mu\nu}=\epsilon_{(\mu}\tilde \epsilon_{\nu)}\,\e^{ik\cdot X}$, $B$-field $B_{\mu\nu}=\epsilon_{[\mu}\tilde \epsilon_{\nu]}\,\e^{ik\cdot X}$ and dilaton $\phi= \epsilon\cdot\tilde \epsilon\, \e^{ik\cdot X}$. These fields form the NS sector of 10d supergravity but make sense in all dimensions and have become known as \emph{fat gravity}.

The array of integrated vertex operators given by \eqref{int-V-ops} are all of the form $w= v^\sL v^\sR$ where $v^\sL$ and $v^\sR$ are 1-forms on the worldsheet, either of the form $t\cdot j$ or $\epsilon\cdot P +  k\cdot \Psi\, \epsilon\cdot \Psi $.  This decomposition gives an elegant microscopic formulation of the double copy with the interchangeability of the two types of operator building up from biadjoint scalars as the zero'th copy to Yang-Mills and then gravity.  
We note that the biadjoint scalar model also contains gravitational and gauge sectors and there is also a gravitational sector in the heterotic model which we briefly discuss in \cref{fdev}.

\paragraph{Fixed vertex operators.} In these models, non-trivial correlators also require three fixed vertex operators, related to the integrated vertex operators above via moduli insertions.  These operators are inserted at three arbitrarily chosen fixed points on $\Sigma$ without integration, and saturate the ghost zero modes in the path-integral.  Much of this arises in the same way as conventional string theory.
For the three models above these come in  the form 
\begin{equation}
V= c\tilde c\,  w \, \e^{ik\cdot X}\, ,
\end{equation}
The lack of integration is associated to fixing residual worldsheet diffeomorphism freedom. 
Setting the worldsheet gravity $e=0$ fixes  the coordinates up to the three-dimensional  group of M\"{o}bius transformations PSL$(2,\C)$. This is a non-compact integration in the path integral, handled via the  Faddeev-Popov procedure by fixing the insertion points of three vertex operators, $(\sigma_1,\sigma_2,\sigma_3)$ to e.g. $(0,1,\infty)$.
The vertex operators remain $Q$-invariant despite the lack  of integration due to the ghosts $c$; in the path integral this amounts to only quotienting by gauge transformations that vanish at these fixed insertions. Since the fermionic ghosts $c\in \Omega^0(T_\Sigma)$ have three zero modes on the sphere (see \ref{sec:CFT}), tree-level correlators must include three fixed vertex operators to give non-trivial Berezinian integrals.\footnote{More precisely $n_c-n_b=3$ must match the zero-mode count, where $n_{b,c}$ are the numbers of respective ghost insertions. The $b$-ghost insertions can arise from moduli insertions.} The ghost correlation function provides the needed Faddeev-Popov determinant associated to the gauge fixing. 

The  most novel part in these models, compared to conventional strings, is the descent associated to the gauge field $\tilde e$ that imposes  the $P^2=0$ constraint.  In this case, the residual gauge freedom amounts to adding $\alpha P$ to $X$ where $\alpha$ is a holomorphic section of $T_\Sigma$.  When $\Sigma=\CP^1$, this is 3-dimensional, and is fixed by fixing the values of $X(\sigma_i)$ at three values of $\sigma$ on the corresponding point of the geodesic.  Descent is given  by the connecting homomorphism $\delta$ described in \eqref{conn-hom} that implements the Penrose transform from space-time fields to their corresponding cohomology classes on ambitwistor space.

\paragraph{Pictures.} The same procedure for the fermionic symmetries associated constraint $P\cdot \Psi$ or $P\cdot \tilde\Psi$ leads to vertex operators (both fixed and integrated) in different pictures. We denote this via a superscript $(p)$, with $p=-1$ or $p=0$,
\begin{align}
& V^{(p)}= c\tilde c\,  w^{(p)} \, \e^{ik\cdot X}\, ,
&&\cV^{(p)}=\int w^{(p)}\, \bar{\delta}(k\cdot P)\,\e^{ik\cdot x}\, , 
\end{align}
and take again $w^{(p)}=v^{(p)\sL}\, v^{(p)\sR}$ in line with the double copy, where
\begin{align}
 & v^{(-1)}= \delta(\gamma)\,\epsilon\cdot \Psi
 && v^{(0)}=\epsilon\cdot P + k\cdot \Psi\, \epsilon\cdot \Psi\,.
\end{align} 
Vertex operators in different pictures are related via so-called picture changing operators $\Upsilon = \delta(\beta)\, P\cdot \Psi$ (and similarly $\tilde\Upsilon= \delta(\tilde\beta)\, P\cdot \tilde\Psi$);
\begin{equation}\label{eq:descent}
	V^{(0)}(\sigma) = \mathop{\mathrm{lim}}_{z\rightarrow \sigma} \; \Upsilon(z) \tilde\Upsilon(z)\; V^{(-1)}(\sigma)\,.
\end{equation}
Note that the vertex operators of \eqref{int-V-ops} are thus given in the picture $p=0$; with $v=v^{(0)}$.
Since there are two zero modes for each bosonic ghost $\gamma$, tree-level correlators must contain exactly two vertex operators with picture number $p=-1$.\textsuperscript{\ref{fn:PCOs}}

\subsection{Amplitudes}
In general amplitudes are obtained as correlation functions of vertex operators with sufficient fixed vertex operators to precisely saturate the zero-modes of the ghost fields;
\begin{equation}
\cA(1,\ldots n)=\left\langle V_1V_2V_3\cV_4\ldots \cV_n\right\rangle\, .
\end{equation} 
In the supersymmetric cases, two of the fixed vertex operators of the form $V^{(-1)}$ and one of type $V^{(0)}$.  Our first task is to see that the path-integral that defines this correlation function localizes onto the solutions to the scattering equations.

In all cases, the computation of the correlator involves a worldsheet correlation function of the $w$'s of \eqref{int-V-ops}
which we denote by $\langle W \rangle_\Sigma$, as well as the correlators of the $c$ and $\tilde c$ ghosts denoted by $\langle C\tilde C\rangle_\Sigma$. We will return shortly to the task of evaluating these correlators. The $PX$-correlator is universal to all models, and we consider it first. Fortunately, we can sidestep the problem of evaluating the contractions between the $X$'s in the  $\e^{ik\cdot X}$ factors and $P$'s in the vertex operators by taking the $\e^{ik\cdot X}$ factors into the action in the path-integral, and treating them as sources in the $PX$-action.  
After gauge fixing this action, the full correlation function is then given by the path integral
\begin{equation}
\left\langle V_1\ldots \cV_n\right\rangle = \int D[X,P]\,\int \prod_{i=4}^n \bar\delta \big(k_i\cdot P(\sigma_i)\big)\, d\sigma_i \; \, \langle C\tilde C\rangle_\Sigma\, \langle W \rangle_\Sigma \; \,e^{S_{\mathrm{eff}}}\, ,
\end{equation}
and we can write the (effective) action with the vertex operator sources as 
\begin{equation*}
S_{\mathrm{eff}} = 
\frac{1}{2\pi} \int_\Sigma P\cdot \dbar X + \sum_{i=1}^n i k_i\cdot X(\sigma_i)
=\frac{1}{2\pi} \int_\Sigma \left( P\cdot \dbar X + \sum_{i=1}^n k_i\cdot X(\sigma))\bar\delta(\sigma-\sigma_i)\, d\sigma\right) \, . 
\end{equation*}
Since the action is now linear in $X$, and there is no further $X$-dependence in the path-integral, we can integrate out the $X$-field. Its zero-modes provide $d$ momentum-conserving delta-functions, while the non-zero-modes localize the $P$ path-integral onto the solution to  the equations of motion of this action:
\begin{equation}\label{eq:EoM_P}
\dbar P_\mu=\sum_{i=1}^n ik_{i\mu}\, \bar\delta(\sigma-\sigma_i)\, .
\end{equation}
On the sphere, these have a unique solution given by
\begin{equation}
P_\mu(\sigma) = \sum_i \frac{k_{i\mu}}{\sigma-\sigma_i}\, .
\end{equation}
This solution can be substituted into the delta-functions (and into the $w$'s) yielding 
\begin{equation}
\bar \delta\big(k_i\cdot P(\sigma_i)\big)=\bar \delta (\cE_i)\, , \qquad \cE_i:= \sum_j \frac{k_i\cdot k_j}{\sigma_{ij}}\, , \qquad \sigma_{ij}=\sigma_i-\sigma_j\, .
\end{equation}
We now see that the delta-functions impose the \emph{scattering equations} $\cE_i=0$.  
Thus the path integral localizes to 
\begin{equation}
\left\langle V_1\ldots \cV_n\right\rangle =  \delta^{d}\Big(\sum_{i=1}^n k_i\Big)\; 
\int \prod_{i=4}^n \bar\delta \big(k_i\cdot P(\sigma_i)\big)\, d\sigma_i \;\, \langle C\tilde C\rangle_\Sigma\, \langle W \rangle_\Sigma \, . 
\end{equation}
The correlator of the three $c$'s and $\tilde c$'s is elementary and gives a numerator factor of $(\sigma_{12}\sigma_{23}\sigma_{31})^2$.  With this we  define the CHY measure
\begin{equation}
  d\mu_n^\CHY := (\sigma_{12}\sigma_{23}\sigma_{31})^2 \prod_{i=4}^n \bar\delta \big(k_i\cdot P(\sigma_i)\big)\, d\sigma_i= \frac{\prod_{i=1}^n\bar\delta(\cE_i)\, d\sigma_i}{\mathrm{vol} \; \mathrm{PSL}(2,\mathbb{C})\times \C^3}\, .
\end{equation}
In the second equality here we have identified one factor of $\sigma_{12}\sigma_{23}\sigma_{31}$ as the Faddeev-Popov determinant for the action of M\"{o}bius transformations on the sphere with $n$ points, and the second factor as that for the action of the residual gauge symmetries \eqref{gauge} associated to translations along the lightray.
This is now a measure on the moduli space $\mathfrak{M}_{0,n}$, the moduli space of $n$ points in $\CP^1$ modulo M\"{o}bius transformations.

To finish the correlation function computation we need to evaluate $\langle W\rangle_\Sigma$.  Because of the construction of the $w$'s as $w=v^\sL v^\sR$, where $v^\sL$ and $v^\sR$ are constructed from independent worldsheet matter systems, this correlation function naturally factorizes as
\begin{equation}
 \langle W\rangle_\Sigma= \cI^\sL \cI^\sR\,,
\end{equation}
where $\cI^\sL$, $\cI^\sR$ are respectively  the correlators of the $v^\sL$'s and $v^\sR$'s.  This gives the final amplitude formula as
\begin{equation}\label{eq:CHY}
 \cA_n^\CHY=\int d\mu_n^\CHY\,\mathcal{I}_n\,,
 \qquad \quad  \cI_n:= \cI_n^\sL \cI_n^\sR.
\end{equation}
All ambitwistor strings give formul{\ae} of this CHY form with $\cI^\sL$, $\cI^\sR$ given as follows.

For the current algebra $S_C$ this correlation function is standard, breaking up into single-trace and multi-trace terms.  The single-trace terms are a sum over permutations $\alpha\in S_n$  of  $\tr (t_{\alpha(1)}\ldots t_{\alpha(n)})\, \Pt(\alpha)$ where the $\Pt(\alpha) $   are \emph{Parke-Taylor} factors defined by  
\begin{equation}
\Pt(\alpha)=\prod_{i=1}^n \frac{1}{\sigma_{\alpha(i) \alpha(j)}}\, .
\end{equation}
The multi-trace terms are also part of the field theory defined by the ambitwistor string, and can be interpreted as tree amplitudes with one of the scalars of the corresponding gravity theory running along an internal propagator  \cite{Witten:2003nn, Witten:2005px}.

The most interesting ingredient is the correlator arising from the $v^\sL$'s when  $S^\sL=S_\Psi$ leading to the CHY Pfaffian, defined via a skew symmetric the $2n\times 2n$ matrix
\begin{equation}
M:=\begin{pmatrix}
A  & C\\ -C^t& B 
\end{pmatrix}\,,
\end{equation}
with components
\begin{equation}
A_{ij}:=\frac{k_i\cdot k_j}{\sigma_{ij}} \, , \qquad B_{ij}:= \frac{\epsilon_i\cdot \epsilon_j}{\sigma_{ij}}\, ,
\qquad
C_{ij}:= \begin{cases}
\frac{\epsilon_j\cdot k_i}{\sigma_{ij}}\, ,& i\neq j\, ,\\
-\sum_k \frac{\epsilon_i\cdot k_k}{\sigma_{ik}}\, ,& i=j\, .
\end{cases}
\end{equation}
The matrix has a 2-dimensional kernel on the support of the scattering equations given by the row vectors $(1,\dots,1\, | \, 0,\dots, 0)$ and $(\sigma_1,\dots,\sigma_n\, | \,0,\dots, 0)$. This allows us to define a reduced Pfaffian  by 
\begin{equation}\label{Pfaffian}
\Pf'(M)=\frac1{\sigma_{12}}\Pf(M^{12}),.
\end{equation}
where $M^{12}$ is $M$ with the first two rows and columns deleted.  Importantly, this reduced Pfaffian is permutation invariant.
We now have the main statement that the correlation function of the $v^\sL$'s from $S_\Psi$ is given as
\begin{equation}
\langle v^0_1v^0_2 v_3\ldots v_n\rangle=\Pf'(M).
\end{equation}
With these ingredients, we have now arrived at the original three main CHY formul{\ae},
\begin{equation}\label{integrands}
\cI_n=\cI^\sL \cI^\sR=\begin{cases}  \Pt(\alpha)\,\Pt(\beta)\, , & \text{Biadjoint scalar}\\
\Pt(\alpha)\,\Pf'(M)\, , &\text{Yang-Mills theory}\\
\Pf'(M)\,\Pf'(\tilde M)\, , &\text{NS gravity}.
\end{cases}
\end{equation}


\subsection{Proof of the CHY \texorpdfstring{formul{\ae}}{formulae}}\label{sec:proof}
The CHY formul{\ae} (\ref{eq:CHY}, \ref{integrands}) are strikingly compact, valid for all multiplicity  and all dimensions, with a tantalizing worldsheet origin.  They are quite remote from standard formulations of field theory scattering amplitudes; so how do we know they correctly describe amplitudes? A straightforward sanity check verifies that  they give the correct 3- and 4-particle amplitudes, which is already nontrivial at four points. 
In this section, we give a full proof for the ambitwistor string correlators $\cA_n^\CHY$ for any number of external particles. Along the way, we will gain a better understanding of the role the scattering equations play for massless amplitudes, and explore how they relate geometric factorization in the moduli space $\mathfrak{M}_{0,n}$ to kinematic factorization of the scattered particles which occurs when partial sums of the momenta become null.

The  proof is based on the Britto-Cachazo-Feng-Witten (BCFW) recursion relation for scattering amplitudes \cite{Britto:2004ap, Britto:2005fq, Cachazo:2005ca, Bedford:2005yy}, reviewed in chapter 1 \cite{chapter1}. Employed constructively, the BCFW recursion allows us to build the full tree-level S-matrix recursively from lower point amplitudes, using the three-particle amplitudes as seeds. Thus, BCFW also guarantees that any proposed expression satisfying the recursion relation,  with the correct three-point seed amplitudes, is a representation of the S-matrix. This means that we can prove the CHY formul{\ae} for Yang-Mills and gravity by 
verifying that they obey the assumptions that lead to the BCFW recursion relations.

\paragraph{BCFW recursion. }
The  on-shell recursion relations exploit elementary complex analysis and knowledge of singularities of the amplitudes. The poles occur precisely at \emph{factorization channels} of the amplitude when a partial sum of the external momenta becomes null so that some intermediate propagator becomes singular. If the theory is both \emph{local} and \emph{unitary}, then these poles are all simple with residues given by  the sum of products of two tree amplitudes 
\begin{equation}\label{eq:fact_ampl}
 \lim_{K^2_I\rightarrow 0}K^2_I\,\cA_n =\sum_r \cA_{n_I+1}\left(K_I,r\right)\,\cA_{n_{\bar I}+1}\left(-K_I,r\right)\,,
\end{equation}
where $K_I=\sum_{i\in I} k_i$ is the partial sum of momenta of the particles $i\in I$, and the sum is over polarization states $r$ that can run in the propagator; here we have denoted the sets of external particles in each of the subamplitudes by $I$ and $\bar I$, with multiplicities $n_I=|I|$ and $n_I+n_{\bar I}=n$.
This is the content of the \emph{optical theorem}, here restricted to tree-level amplitudes (stripped of their momentum-conserving delta-functions).
It ensures that tree-level amplitudes 
are meromorphic functions of the external momenta, with only simple poles.

This allows us to harness the power of complex analysis in which one can reconstruct a holomorphic function on the Riemann sphere from its residues.  To exploit this idea, we choose a complex one-parameter deformation of the external momenta, 
\begin{equation}\label{eq:shift}
 k_1\rightarrow \hat{k}_1(z)=k_1+zq\,,\qquad k_n\rightarrow \hat{k}_n(z)=k_n-zq\,,
\end{equation}
where $z$ is a complex variable and $q_\mu$ is a reference vector satisfying $q^2=q\cdot k_1=q\cdot k_n=0$ such that all external particles remain on-shell. \footnote{For  particles transforming in non-trivial representations of the little group, the polarization vectors have to be shifted as well, which is best seen in covariant gauge, see e.g. \cite{Arkani-Hamed:2008bsc} for details.}
Then Cauchy's residue theorem allows us 
to express the original undeformed  amplitude as the sum over all other residues, including a boundary term $\mathcal{B}_{\infty}$ from a potential residue as $z\rightarrow \infty$,
\begin{equation}\label{eq:BCFW}
 \cA_n=\frac{1}{2i\pi}\oint_{z=0} \frac{\cA_n(z)}{z} = \sum_{I,r_I} \frac{1}{K_I^2}\,\cA_{n_I+1}(z_I,r_I)\,\cA_{n_{\bar I}+1}(z_I,r_I)+\mathcal{B}_{\infty}\,.
\end{equation}
If for some good choices of $q$, the boundary term vanishes, $\mathcal{B}_{\infty}=0$, a theory is \emph{on-shell constructible}.
The remaining residues  away from infinity correspond to singular kinematic configurations, where the optical theorem guarantees that the amplitude factorises into a product of on-shell lower-particle amplitudes, giving the BCFW recursion for scattering amplitudes.
Thus the tree S-matrix can  be built from three-point amplitudes. This is the case for theories such as Yang-Mills, gravity \cite{ArkaniHamed:2008yf} or indeed any 4d renormalizable QFT  \cite{Cohen:2010mi}; see  the review chapter 1 \cite{chapter1}, or  \cite{Elvang:2013cua, Elvang:2015rqa, Cheung:2017pzi, Feng:2011np}.

\paragraph{Factorization proof of the CHY representation}
The CHY formul{\ae} will satisfy  the BCFW recursion \eqref{eq:BCFW} if we can prove that it satisfies factorization \eqref{eq:fact_ampl} and $\cB_\infty=0$.
The solution to the recursion is unique, given appropriate three-point seeds, and so we can deduce that the CHY formul{\ae} give  valid representations of the amplitude. 
We will use this strategy following ref.~\cite{Dolan:2013isa}.

\begin{figure}[ht]
\begin{center}
 \includegraphics[width=6cm]{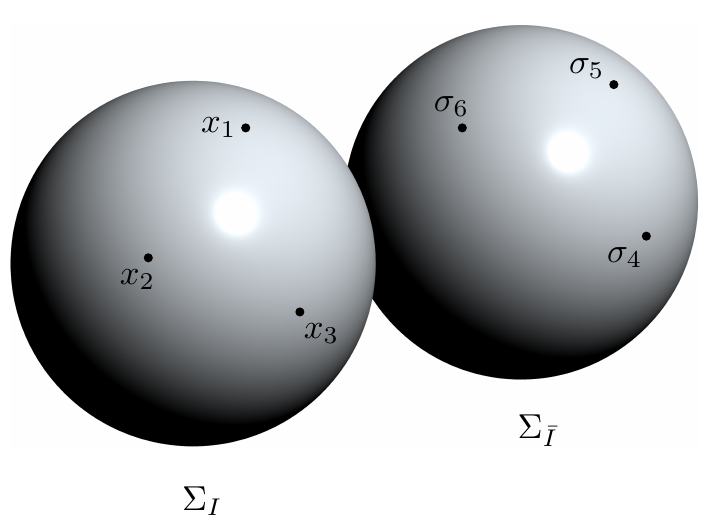}
 \caption{Boundary divisor $\mathfrak{M}_{0,n_I+1}\times\mathfrak{M}_{0,n_{\bar I}+1}\subset \partial\widehat{\mathfrak{M}}_{0,n}$ of the moduli space, corresponding to a pair of marked spheres $\Sigma_I$ and $\Sigma_{\bar I}$. We parametrize the spheres by $x\in \Sigma_I$, and $\sigma\in \Sigma_{\bar I}$, with a nodal point $\sigma_I\in \Sigma_{\bar I}$ and $x_I=\infty\in\Sigma_I$, and subject to $\sigma=\sigma_I+\varepsilon x$.
 Depicted is the case for $n=6$, with $I=\{1,2,3\}$ and $\bar I = \{4,5,6\}$.}
 \label{fig:sphere_degen}
\end{center}
\end{figure}

Recalling the general structure of $\cA_n^\CHY$, we can see  that the formul{\ae} only have poles  when there are residues where a subset $I$ of the marked points collide,
 \begin{equation}\label{eq:fact_sigma}
 \sigma_i=\sigma_I+\varepsilon x_i \qquad\mathrm{for }\;i\in I\;\;\; \mathrm{ and }\;\;\;\varepsilon\rightarrow 0 \,.
\end{equation}
This parametrizes a boundary $\partial\widehat{\mathfrak{M}}_{0,n}$ of the (Deligne-Mumford compactified  \cite{DeligneMumford, Witten:2012bh}) moduli space, corresponding to a separating degeneration of the worldsheet  into a pair of spheres  connected by a node $\sigma_I$, see  \cref{fig:sphere_degen}. Thus, the CHY formula only  generates  poles when solutions to the scattering equations approach the boundary of the moduli space. We now show that this can only happen if the corresponding $K_I$ is null.

The scattering equations relate these boundaries $\partial\widehat{\mathfrak{M}}_{0,n}$ to the factorization channels of the amplitude as follows.  The key observation is that the scattering equations descend naturally onto the two component  spheres $\Sigma_I$ and $\Sigma_{\bar I},$
\begin{equation}\label{eq:degen_SE}
 \mathcal{E}_i=
 \begin{cases}
 \frac{1}{\varepsilon}\mathcal{E}^{(I)}_i 
   & i\in I \\
   \mathcal{E}^{(\bar I)}_i
    & i\in \bar I\,,
 \end{cases}
\end{equation}
where the equality holds to leading order in the degeneration parameter $\varepsilon$, and where $\mathcal{E}^{(I)}$ and $\mathcal{E}^{(\bar I)}$ are the scattering equations on $\Sigma_I$ and $\Sigma_{\bar I}$ respectively,
\begin{equation}
 \mathcal{E}^{(I)}_i   := \sum_{j\in I}\frac{k_i\cdot k_j}{x_{ij}} \,,
 \qquad\qquad 
 \mathcal{E}^{(\bar I)}_i
   :=\sum_{p\in \bar I}\frac{k_i\cdot k_p}{\sigma_{ip}} + \frac{k_i\cdot K_I}{\sigma_{iI}}\,,
\end{equation}
with $K_I = \sum_{i\in I} k_i$ the momentum of the internal propagator.\footnote{In these expressions, we implicitly gauge-fixed the nodal point (from the perspective of $\Sigma_I$) to $x_I=\infty$.} 
Under the degeneration \eqref{eq:fact_sigma} and on the support of the full scattering equations, both of these sets of scattering equations are of $O(\varepsilon)$. This guarantees that also the momentum $K_I^2=O(\varepsilon)$ of the propagator connecting the two sub-amplitudes is on-shell because
\begin{equation}
 K_I^2=\frac{1}{2}\sum_{i,j\in I}k_i\cdot k_j=\sum_{\substack{i,j\in I\\ i\neq j}} \frac{x_ik_i\cdot  k_j}{ x_i-x_j}= \sum_{i\in I} x_i\mathcal{E}_i^{(I)}=\mathcal{O}(\varepsilon)\,.
\end{equation}
The scattering equations thus map the boundary of the moduli space  to factorization channels, with singular kinematic configurations. This property 
lies at the heart of the CHY formalism, and it is the reason the scattering equations are universal to massless theories, elegantly   encoding massless propagators.  The role of the integrand on the other hand is to determine which poles occur in the amplitude, and to provide the correct numerator structure -- all information specific to a given theory.  

The essence of factorization in this framework now reduces to a straightforward scaling argument in the degeneration parameter $\varepsilon$ to determine the degrees of and residues  at the poles arising for the factorization channels. Consider first the measure $d\mu_n^\CHY$: the scattering equations and factors $d\sigma_i$ descend naturally to $\Sigma_{I,\bar I}$, each providing one power of $\varepsilon$ for particles $i\in I$. The only subtlety stems from M\"{o}bius invariance on $\Sigma_I$, which allows us to fix  $x_m=1$ for some $m\in I$, such that $d\sigma_m=d\varepsilon$. Moreover, on the support of the remaining scattering equations, one equation enforces the constraint $K_I^2=\mathcal{O}(\varepsilon)$. Combining all these factors, we see that the measure factorizes neatly into
 \begin{equation}\label{eq:fact_SE}
  d \mu_n=\varepsilon^{2\left(n_I-1\right)}
  \frac{d\varepsilon}{\varepsilon}  \,\bar{\delta}\left(K_I^2+\varepsilon\mathcal{F}\right)\,d\mu_{n_I+1}\, d\mu_{n_{\bar I}+1}\,.
\end{equation}
Here, the measure $d\mu_{n_I+1}$ is the natural CHY measure for the sphere $\Sigma_I$; with $n_I$ marked points corresponding to the external particles and the additional nodal point associated with the internal propagator. As we saw above, the delta-function restricts $K_I^2$ to vanish to order $\varepsilon$, thereby restricting to singular kinematic configurations. The factor $\mathcal{F}$ denotes a function of the marked points and kinematics, and  will drop out of the final formula for a unitary theory with simple poles. 

At this point, we can see that the integrand $ \mathcal{I}_n$  for a theory respecting the factorization property \eqref{eq:fact_ampl} must split into respective integrands on each sphere;
\begin{equation}\label{eq:fact_int}
 \mathcal{I}_n=\varepsilon^{-2\left(n_I-1\right)}\, \sum_{\mathrm{states}}\mathcal{I}_{n_I+1}\,\mathcal{I}_{n_{\bar I}+1}
 \,.
\end{equation}
We can phrase this equivalently as a condition on the `half-integrands' $\cI^{L/R}$, now with a scaling factor of $\varepsilon^{-\left(n_I-1\right)}$.
This property can easily be verified for the Parke-Taylor factors, and was proven for the reduced Pfaffian $\pf{}'M$ in \cite{Dolan:2013isa}. For the bi-adjoint scalar, Yang-Mills theory and gravity we can then combine the behaviour of  the  measure  \eqref{eq:fact_SE} and the integrands \eqref{eq:fact_int}, 
\begin{equation}
\cA_n=\sum_{\mathrm{states}} \int  \frac{d\varepsilon}{\varepsilon} \,\bar{\delta}\left(K_I^2+\varepsilon\mathcal{F}\right)\, \left(d\mu_{n_I+1}\mathcal{I}_{n_I+1}\right)\left(d\mu_{n_{\bar I}+1}\mathcal{I}_{n_{\bar I}+1}\right)\,,
\end{equation}
to see that the amplitude exhibits factorization properties in line with the optical theorem and the BCFW recursion,
\begin{equation}
\cA_n=\frac{1}{K_I^2}\sum_{\mathrm{states}}\cA_{n_I+1}\cA_{n_{\bar I}+1}\,.
\end{equation}
The pole structure of the amplitude is thus determined by an interplay  between the integrand and the measure including the scattering equations: while the measure guarantees the correct form for \emph{all} possible poles, the integrand selects a subset of the poles  suitable for the theory, and determines the residue on the pole via the factorized integrands $\mathcal{I}_{n_I+1}$ and $\mathcal{I}_{n_{\bar I}+1}$.  For example, for a Parke-Taylor factor, we only find a pole when the subset $I$ is connected in the cyclic ordering of the Parke-Taylor.

\paragraph{The boundary term:} For both gravity and Yang-Mills theory, the absence of the boundary term can be verified in covariant gauge \cite{Arkani-Hamed:2008bsc}  with a careful choice of shift vector $q=\epsilon_1$.  Considering the shift \eqref{eq:shift} in the limit $z\rightarrow \infty$, the two shifted momenta become to leading order $k_{1,n}=\pm zq$, so that the amplitude has an interpretation as a  hard light-like particle scattering propagating in a soft background. On the support of the scattering equations, the dominant contribution to the amplitude then stems from the boundary of moduli space where $\sigma_{n}-\sigma_1=\varepsilon$, with the degeneration parameter $\varepsilon$ scaling as $z^{-1}$. The calculation for the overall scaling behaviour in $z$ is then lengthy but straightforward: we find that the measure scales as $d\mu^\CHY\sim z^{-2}$, whereas the integrands\footnote{if the shifted particles are non-adjacent in the planar ordering $\alpha$,  this improves to $\mathrm{PT}(\alpha)\sim z^0$} behave as  $\mathrm{PT}\sim z$  and $\pf{}'M\sim z^0$. For both gauge theory and gravity, this ensures that the boundary term vanishes,
\begin{equation}
 \mathcal{B}_\infty = \lim_{z\rightarrow\infty}\cA_n^\CHY = 0\,.
\end{equation}
This concludes our proof of the CHY representation for Yang-Mills theory and gravity.

\subsection{Discussion}\label{fdev}
The ambitwistor string theories of this section are not simply vehicles to arrive at CHY formul{\ae}, but contain much more information. In \S\ref{Loops} we will see that they provide a stepping stone to loop integrands.  Here we mention other features. Theories often contain extra vertex operators beyond those originally envisaged.  These  extend the tree formul{\ae} to amplitudes of more elaborate theories combining gravity, gauge theory and scalars.  Furthermore,  different forms of worldsheet matter lead to different theories, and different representations of ambitwistor space give  different amplitude representations. We also make some brief remarks on connections with the colour-kinematic duality.
.

 Although the model $S_{\mathrm{BAS}}$ leads directly to the CHY  biadjoint scalar amplitude formula above, it also contains gauge theory and gravity vertex operators that lead to formul{\ae} for gauge and gravity amplitudes.  These were initially hard to interpret with the simplest having $\cI^\sR=\prod_i \epsilon_i\cdot P(\sigma_i)$ for the gauge theory, and doubled for the gravity theory.  They are now understood to be parts of a 4th order gauge theory that is conformally invariant in 6d as described in \cite{Johansson:2017srf} and a 6th order gravity theory whose linearization is given in \cite{Berkovits:2018jvm}, see \cite{Geyer:2020iwz} for discussion of these theories in 6d.

Similarly $S_{\mathrm{het}}$ contains vertex operators for a 4th order gravity that is conformally invariant in 4d and is thought to agree \cite{Johansson:2017srf} in 4d with that found by Berkovits and Witten  \cite{Berkovits:2004jj}, but extends to all dimensions \cite{Azevedo:2017lkz}.  What is remarkable in both these examples is that the ambitwistor-string  models are able to generate amplitudes for complete theories including gauge and gravity sectors, albeit ones that are pathological with higher-order equations of motion.

Like the conventional string, the type II  gravity models  $S_{\mathrm{II}}$ require a GSO projection to project out unwanted states; we only gave vertex operators consistent with the GSO projection that are even under  $(\gamma, \Psi)\rightarrow (-\gamma,-\Psi)$ and similarly for  $(\tilde \gamma,\tilde{\Psi})$.  These models also admit a Ramond sector constructed in the usual way from the $\Psi$ spin-field, with vertex operators for   NS-R and R-R sectors as in the conventional string in addition to the NS-NS vertex operators introduced above. They require the GSO projection to be applied independently to the `tilded' and `untilded' states to yield the 10d type IIA and IIB supergravity theories \cite{Adamo:2013tja}. In principle, correlators give amplitude formul{\ae} for all sectors and any multiplicity, but  Ramond-sector correlators are  hard to compute explicitly  beyond three and four points.  Again we see that the ambitwistor-string model naturally completes the NS-NS-sector CHY-formul{\ae} to the well-known type-II supergravity theories.

Soon after their original formul{\ae}, CHY introduced expressions for amplitudes in many more theories \cite{Cachazo:2014nsa, Cachazo:2014xea}, going beyond Einstein (E), to include Born-Infeld (BI), Dirac-Born-Infeld (DBI), nonlinear sigma models (NLSM), Einstein-Maxwell (EM), Einstein-Yang-Mills (EYM),  Yang-Mills Scalar (YMS) and even galileons.   Further forms of worldsheet matter can be introduced to yield models that generate these formul{\ae} as in  table \ref{models}. Here for example $S_{\Psi_1,\Psi_2}$ is an $N=2$ version of the worldsheet superalgebra \cite{Ohmori:2015sha} that lives discussed above, and we refer to \cite{Casali:2015vta} for full details of all the models.

\begin{table}[ht]{
\begin{tabular}{|c||l|l|l|l|l|}
  \hline
  \diagbox{$S^\sL$}{$S^\sR$}& $S_\Psi$ & $S_{\Psi_1,\Psi_2}$ & $S_{\rho,\Psi}^{(\tilde{m})}$ & $S_{\mathrm{YM},\Psi}^{(\tilde{N})}$ & $S_{\mathrm{YM}}^{(\tilde{N})}$\\ \hhline{|=||=|=|=|=|=|}
  $S_\Psi$ & E &  & &  & \\ \hline 
  $S_{\Psi_1,\Psi_2}$ & BI & Galileon &  &  & \\ \hline  \Tstrut
  $S_{\rho,\Psi}^{(m)}$ & EM$\big|_{\scalebox{0.65}{U$(1)^{m}$}}\!$ & DBI & EMS$\big|_{\scalebox{0.65}{U$(1)^{m}\otimes $U$(1)^{\tilde{m}}$}}$ &  & \\ \hline \Tstrut
  $S_{\mathrm{YM},\Psi}^{(N)}$ & EYM & extended DBI & EYMS$\big|_{\scalebox{0.65}{SU$(N)\otimes $U$(1)^{\tilde{m}}$}}$ & EYMS$\big|_{\scalebox{0.65}{SU$(N)\otimes $SU$(\tilde{N})$}}\!$ & \\ \hline   \Tstrut
  $S_{\mathrm{YM}}^{(N)}$ & YM &  NLSM & YMS$\big|_{\scalebox{0.65}{SU$(N)\otimes$U$(1)^{\tilde{m}}$}}$ & gen.\! YMS$\big|_{\scalebox{0.65}{SU$(N)\otimes $SU$(\tilde{N})$}}\!$ & BS
  \\ \hline 
 \end{tabular}}
 \caption{Theories arising from the different choices of matter models.} \label{models}
\end{table}

Potentially the most remarkable of these models would be that for Einstein-Yang-Mills, critical in 10d.  However, it has twice as many gluon  vertex operators as appropriate for conventional Yang-Mills theory and describes amplitudes for a theory with action $\int \tr (B \wedge D_A {}^*F_A)$ where $A$ is a standard gauge  field, and $B$ a Lie-algebra valued 1-form serving as a Lagrange multiplier for the Yang-Mills equations on $A$.

The proof of the novel CHY formul{\ae} arising from  the  massless theories listed in  \cref{models} is not immediate: the factorization arguments extend to these models, but   the BCFW shift  is not directly applicable  because of higher powers of momentum dependence in the vertices of many of these theories.  However, different on-shell recursion relations generalizing the BCFW construction  \cite{Cheung:2015ota, Luo:2015tat}, can be used instead. The arguments given above then extend to the whole zoo of theories with CHY-representations with only minor adjustments.

Shortly after the original ambitwistor string \cite{Mason:2013sva},  a pure spinor analogue was  introduced in \cite{Berkovits:2013xba}, and its corresponding amplitude formul{\ae} verified in \cite{Gomez:2013wza}. Models based  on  the Green-Schwarz worldline model appears in \cite{Chandia:2016dwr}, and
progress towards 11d models from their worldline counterparts are treated  in \cite{Berkovits:2019szu,Guillen:2020mmd}.

\paragraph{Color-kinematics duality.}
The double copy \cite{Bern:2010ue} is built into the structure of the CHY formul{\ae} and the ambitwistor string vertex operators.  This is much as in the conventional string, where vertex operators are  a product of left and right moving parts; however, here  in the ambitwistor string, both are holomorphic on the worldsheet. 
In a field theory framework, the double copy  is build on  colour-kinematics duality \cite{Bern:2008qj}: it asserts that Yang-Mills amplitudes can be expressed as a sum over trivalent graphs $\Gamma$ of the form
\begin{equation}
\cA =\sum_{\Gamma}\frac{N_\Gamma\, c_\Gamma}{D_\Gamma}\,,
\end{equation}
where $c_\Gamma$ are the colour factors associated with the graph with vertices determined by the Lie algebra structure constants, $D_\Gamma$ the propagator denominators associated to $\Gamma$ and $N_\Gamma$ the \emph{kinematic numerators} depending on the polarizations and momenta.  If  polynomial in the momenta they are said to be local. Colour-kinematic duality is the assertion that they can be constructed so as to satisfy the same identities as the colour factors $c_\Gamma$ that arise as a consequence of Jacobi identities.  This  implies a double copy in the following form: that by replacing the $c_\Gamma$ by another set of $N_\Gamma$, we obtain Einstein gravity amplitudes or loop integrands \cite{Chapter2}.
Although such local kinematic numerators are known to exist at low loop order, their general theory and underlying kinematic algebra structure remains obscure. The CHY formul{\ae} have provided a powerful tool for construction of such numerators starting with \cite{Cachazo:2013iea} followed by a construction for local numerators  in \cite{Fu:2017uzt} based on the Einstein-Yang-Mills formul{\ae}, see also  \cite{Mizera:2019blq,Edison:2020ehu,He:2021lro,Chapter2} for more recent works with many more references to progress in this very active area.

The structure of the colour factors is best understood in the language of free Lie algebras or Lie polynomials \cite{Frost:2019fjn, Mafra:2020qst, Frost:2020eoa}.  These are embedded in the geometry of the boundary structure of  $\mathfrak{M}_{0,n}$ and play a key role \cite{Frost:2019fjn} in the CHY formula and in the polytope constructions of \cite{Arkani-Hamed:2017jhn}.   See \cite{Frost:2020rjy, Frost:2021qju} for more on numerators in this framework.


\section{The twistor-string and the 4d ambitwistor string}\label{sec:twistor-models}
For the models discussed in the previous section, the $P^2=0$ constraint is gauged in the quantum theory. Nevertheless, the fact that $P^2$ has vanishing OPE with itself\footnote{ Chiral strings for which this is not the case are discussed in \S\ref{chiral}}
 allows this constraint to be solved essentially classically so that the integrated vertex operators are  Penrose transforms of the space-time fields on $\A$ and hence localize on the scattering equations. In this section we discuss models in which the $P^2=0$ constraint is solved explicitly rather than gauged, using spinors classically before quantization. We achieve this by setting
 \begin{equation}
 P_{\alpha\dot\alpha}= \lambda_\alpha\tilde \lambda_{\dot\alpha}\, ,\qquad \qquad \alpha=0,1, \quad\dot\alpha=\dot 0,\dot1\,,
 \end{equation}
where $\lambda_\alpha, \tilde \lambda_{\dot\alpha}$ are two-component spinors  defined up to $(\lambda_\alpha, \tilde \lambda_{\dot\alpha})\rightarrow  (s\lambda_\alpha, s^{-1}\tilde \lambda_{\dot\alpha})$ for $s\neq 0$.

Both the original twistor-string and the 4d ambitwistor-string have the same classical target, the original 4d ambitwistor space, but  choices of twists of line bundles are distinct in the two models (perhaps three models if one includes a `dual-twistor'-string).  They even share some of their vertex operators, and so we present them alongside each other. The 4d ambitwistor-string is framed similarly to RNS type models above and so we present its amplitude formul{\ae} first.

\begin{figure}[ht]
\begin{center}
  \includegraphics[width=7cm]{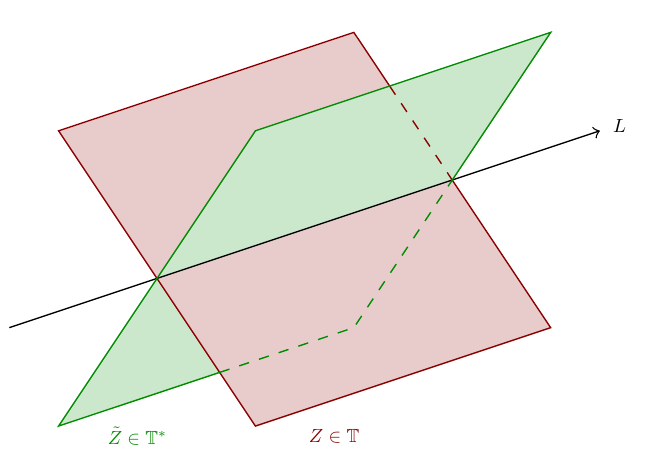}
 \caption{Illustration of ambitwistor space. A fixed twistor $Z\in \P\T$ corresponds to a totally null 2-plane on space-time, known as an $\alpha$-plane; similarly for $\tilde Z\in \P\T^*$ and $\beta$-planes. When $Z\cdot\tilde Z=0$, these planes intersect in a light-ray $L$.}
 \label{fig:twistors}
\end{center}
\end{figure}

The term ambitwistor space arose from the fact that in four dimensions, $\A$ is both the cotangent bundle of chiral  projective twistor space $\P\T$ and its antichiral dual, $\P\T^*$.  
We can supersymmetrize so that twistor space $\T=\C^{4|\cN}$ has coordinates $Z=(\lambda_\alpha,\mu^{\dot\alpha}, \chi^I)$, where 
$\chi^I$ are fermionic coordinates with $I$  an $\cN$-component R-symmetry index.  Similarly we denote a dual twistor by $\tilde Z=(\tilde \lambda_{\dot\alpha},\tilde \mu^{\alpha},\chi_I)\in \T^*$ with the duality defined by  
\begin{equation}
 Z\cdot \tilde Z:=\lambda_\alpha\tilde\mu^\alpha+\mu^{\dot\alpha}\tilde\lambda_{\dot\alpha}+\chi^I\tilde\chi_I\, .
\end{equation}
The original supersymmetric ambitwistor space  of \cite{Witten:1978xx, Isenberg:1978kk}  is a supersymmetric extension of the space of
complex null geodesics which we shall again denote by $\A$ and can be expressed as
\begin{equation}
\A:=\{(Z,\tilde Z)\in \PT\times \PT^*| Z\cdot \tilde Z=0\}/\{ Z\cdot \p_Z-\tilde Z\cdot \p_Z\}\, .
\end{equation}
To see the connection with null geodesics, 
we first introduce the  supertwistor  incidence relations
\begin{equation}
\mu^{\dot \alpha}=(ix^{\alpha\dot\alpha} -\theta^{I\alpha}\tilde\theta^{\dot\alpha}_I)\lambda_\alpha\, ,\qquad  \chi^I=\theta^{I\alpha}\lambda_\alpha\, ,
\end{equation}
which defines an $\alpha$-plane, a totally null self-dual $2|3\cN$-plane in complex Minkowski space $\M^{4|4\cN}$ with coordinates $(x^{\alpha\dot\alpha},\theta^{I\alpha},\tilde\theta^{\dot\alpha}_I)$.  A $\beta$-plane, again a  totally null $2|3\cN$-plane, but now anti-self-dual, is  given by the dual-twistor incidence relations 
\begin{equation}
\tilde \mu^{ \alpha} 
=(-ix^{\alpha\dot\alpha} -\theta^{I\alpha}\tilde\theta^{\dot\alpha}_I)\tilde \lambda_{\dot \alpha}\, ,\qquad \tilde \chi_I=
\tilde \theta_I^{\dot \alpha}\tilde \lambda_{\dot \alpha}\, .
\end{equation}
An $\alpha$-plane and a $\beta$-plane
intersects in a super-null geodesic when $Z\cdot \tilde Z=0$.
Fixing  $(Z,\tilde Z)\in \A$, the coordinates  $(x,\theta,\tilde\theta)$ in super Minkowski space then vary  over a $1|2\cN$-dimensional \emph{super light ray} as illustrated in \cref{fig:twistors}. 
Note that ambitwistor space $\A$ is a phase space with   symplectic potential 
\begin{equation}
\Theta =\frac i2(Z\cdot \rd \tilde Z -  \tilde Z\cdot \rd Z)
=P_{\alpha\dot\alpha} dx^{\alpha\dot\alpha} + \mbox{  fermionic coordinates.
}
\end{equation}
For more details on twistor- and ambitwistor space, as well as the twistor correspondence, we refer the reader to the excellent reviews and textbooks \cite{Adamo:2017qyl, Adamo:2011pv, Huggett:1986fs, Woodhouse:1985id}.

Both the twistor-string and the  4d ambitwistor string consist of worldsheet fields  $(Z,\tilde Z)$ on the  worldsheet Riemann surface $\Sigma$ and taking values  in $\T\times \T^*$ tensored with line bundles on $\Sigma$.  For  the 4d ambitwistor string, $(Z,\tilde Z)$ are  valued also in the spin bundle $K^{\nicefrac{1}{2}}_\Sigma$, whereas for the twistor-string, $Z$ takes values in a line bundle $L\rightarrow \Sigma$ of degree $d\geq 0$ and $\tilde Z$ in $K_\Sigma\otimes L^*$. In both cases the action is based on the symplectic potential with the constraint $Z\cdot \tilde Z=0$ imposed by a Lagrange multiplier $a\in \Omega^{(0,1)}_\Sigma$,
\begin{equation}
S=\frac1{2\pi}\int_\Sigma \tilde Z \cdot \dbar Z-Z\cdot \dbar \tilde Z  +a Z\cdot \tilde Z\, .
\end{equation}
 The gauge symmetry associated with $a$ implements the quotient by $Z\cdot\p_Z- \tilde Z\cdot \p_{\tilde Z}$.  The key distinction\footnote{ Strictly speaking, for the realization $\A=T^*\P\T$ we should base the action on  the symplectic potential $\Theta=i\tilde Z\cdot dZ$; this differs from  that above by the exact form $d(Z\cdot \tilde Z)$ adding an exact term to the stress-energy tensor. } between the two theories at this stage is the assignment of degrees of the line bundles for  $(Z,\tilde Z)$  on the worldsheet.

As before we should include  worldsheet gravity 
by starting with the operator $\bar\p=\bar\p_0+ e\p$.  We gauge fix $e=0=a$ leading to the usual ghost $(b,c)$-system and  ghosts $(u,v)$ for $a$, leading to the BRST operator 
\begin{equation}
Q=\int c T + u\,Z\cdot \tilde Z\, ,
\end{equation}
where $T$ is the worldsheet stress tensor. The potential gauge anomaly vanishes precisely for maximal supersymmetry with $\cN=4$. To have vanishing central charge, we must include additional worldsheet matter, which for super Yang-Mills theory we take to be a current algebra $S_C$ with central charge $\mathfrak{c}=12$.  

 Amplitudes are again obtained as correlation functions of vertex operators.  In the following we just give  integrated vertex operators (they simply differ by factors of $c$ and $u$ from their fixed counterparts), and divide by the volume of $\GL(2,\C)$ in the final formula, understood in the usual Faddeev-Popov sense.

In order to construct vertex operators, we need to encode plane waves on twistor space.  This is achieved by the \emph{Penrose transform}, which realizes massless fields of helicity $h$ as cohomolgy classes $H^1(\P\T, \cO(2h-2))$ and $H^1(\P\T^*, \cO(-2h-2))$  on bosonic twistor space or its dual, and these can be represented as $\dbar$-closed as $(0,1)$ forms of homogeneity $\pm 2h-2$ on $\T$ or $\T^*$ respectively.   For   Maxwell super-momentum eigenstates with bosonic 4-momentum $k_{\alpha\dot\alpha}=\kappa_\alpha\tilde \kappa_{\dot\alpha}$ and fermionic supermomenta $q^I$ or $\tilde q_I$, we have on twistor and dual twistor space respectively
\begin{align}
 a&= \int\frac{\rd s}{s}\bar{\delta}^{2}(\kappa_\alpha-s\lambda_\alpha)\e^{is\left([\mu\,\tilde{\kappa}]+\chi^{I}\tilde{q}_{I}\right)} \quad\in H^1(\PT, \cO)\,,  \label{YM-reps_PT}\\
\tilde a&=\int\frac{\rd s}{s}\bar{\delta}^{2}(\tilde{\kappa}_{\dot\alpha}-s\tilde{\lambda}_{\dot\alpha})\e^{is\left(\la\tilde{\mu}\,\kappa\ra+\tilde{\chi}_{I}q^{I}\right)}\quad\in H^1(\PT^*, \cO)\,. \label{YM-reps}
\end{align}
As before, for a complex variable $z$, $\bar\delta(z)= \dbar( 1/z)$ is a complex double delta-function $(0,1)$-form. These representatives encode the full supermutliplet, and the individual fields can be identified as the coefficients in the expansion in $\chi^I$, each an element of the cohomology class $H^1(\P\T, \cO(2h-2))$ with the appropriate helicity  $h$. For each field, the (dual) twistor representatives  are related to space-time fields by  explicit integral formul{\ae} \cite{Huggett:1986fs}.
Using these  it is straightforward  to check that at $\cN=0$, $a$ and $\tilde {a}$ are representatives for classes in $H^1_{\bar\p}(\P\T)$ and $H^1_{\bar\p}(\P\T^*)$ respectively that generate  Maxwell field  momentum eigenstates $\tilde \kappa_{\dot\alpha}\tilde \kappa_{\dot\beta}\, \e^{ik\cdot x}$ and conjugate.  
These degree-zero cohomology classes 
pull  back  to ambitwistor space $\A$ and combine to give the spin-$1$  plane wave representative \eqref{drep}. For gravity the relationship is more subtle.

\paragraph{Yang-Mills vertex operators.} For both the twistor-string and the 4d ambitwistor-string, the Yang-Mills vertex operators are constructed from the representatives above, multiplied by current-algebra generators $t\cdot j$ as for the `vector' models discussed in the last section.
In the twistor string, vertex operators are built from the supertwistor representatives $a(Z)\in H^1(\P\T, \cO)$ with full $\cN=4$ supersymmetry; for these, the expansion in $\chi^I$ gives the full spin-1  supermultiplet represented on twistor space. In the ambitwistor string, vertex operators can also be constructed from the dual $\tilde a(\tilde Z)\in H^1(\P\T^*, \cO)$, again with maximal supersymmetry. Explicitly,
\begin{align}
\cV'_a=\int_\Sigma
 a_a\,  j\cdot t_{a}\, , \qquad 
 \label{YM'}  
\qquad\widetilde  \cV_a= \int_\Sigma
\tilde  a_a\,  j\cdot t_{a}\, ,
\end{align}
where $a$ is a particle label, $a=1,\dots,n$.
These vertex operators are straightforwardly consistent and $Q$-invariant.  However,
the supermomenta $q^I$ and $\tilde q_I$ are not independent, but are Fourier transforms of each other;
a uniform representation is obtained  by a fermionic Fourier transform on $\cV'_a$, giving
\be\label{YM-vertex}
\cV_a=\int \frac{\rd s_a}{s_a} \,\bar\delta^{2|\cN}(\kappa_a-
s_a\lambda \,|\, q_i-s_i\chi)\,j\cdot t_a , \e^{is_a[\mu\,
  \tilde\kappa_a]}\, ,
\ee
where for a fermionic variable $\chi$, $\delta(\chi)=\chi$.

\subsection{Yang-Mills amplitudes in the 4d ambitwistor string.}
N$^{k-2}$MHV Yang-Mills amplitudes can be obtained as correlation functions of the above vertex operators, taking $k$ from dual twistor space and $n-k$ from twistor space:
\begin{equation}
\cA=\left\la \widetilde \cV_1 \ldots \widetilde \cV_k \cV_{k+1}\ldots  \cV_n\right\ra\, .
\end{equation}
The current algebra correlator gives the usual Parke-Taylor factor (together with some multitrace terms to be discussed later).  As before,   we take the exponentials from the vertex operators into the action to provide source terms 
\begin{equation}
\int_\Sigma  \sum_{i=1}^k  i s_i , (\la \tilde \mu \kappa_i\ra +\tilde \chi\cdot q_i)\bar\delta(\sigma-\sigma_i) + \sum_{p=k+1}^nis_p\, [\mu\,\tilde\kappa_p]\, \bar\delta(\sigma-\sigma_p)\,. 
\end{equation}
With these sources, the equations of motion for $Z$ and $W$ become
\begin{equation}
\bar{\partial} Z=
\sum_{i=1}^{k}s_{i}\left(\kappa_{i},0,q_{i}\right)\,{\bar\delta}\left(\sigma-\sigma_{i}\right), 
\quad \qquad \label{dbar-l}
\bar{\partial} W=
\sum_{p=k+1}^{n}s_p\left(\tilde{\kappa}_{p},0,0\right)\,{\bar\delta}(\sigma-\sigma_{p}) .
\end{equation}
Since $(Z,\tilde Z)$ are worldsheet spinors, the solutions exist and are unique, given by
\begin{equation}\label{eom-soln}
Z(\sigma)
=\sum_{i=1}^{k}\frac{s_{i}\left(\kappa_{i},0,q_{i}\right)}{\sigma-\sigma_{i}}
\, , \qquad\quad
\tilde Z(\sigma)
=\sum_{p=k+1}^{n}\frac{s_{p} \left(\tilde{\kappa}_{p},0,0\right)}{\sigma-\sigma_{p}} \, .
\end{equation}
Thus the correlator localizes on the integrals
\begin{equation}
\label{final-form}
\cA=\int \Pt(\alpha)\, d\mu^{(\mathrm{4d})}_n\, , \qquad 
\end{equation}
where $d\mu^{(\mathrm{4d})}_n$ is the 4d \emph{polarized scattering equations} (explained below) measure:
\begin{equation}
d\mu^{(\mathrm{4d})}_n:=\frac{\prod_{a=1}^n  \rd\sigma_a\, \rd
  s_a /s_a }{\mathrm{vol} \,\GL(2,\C)}   
  \prod_{i=1}^k  \bar\delta^2\big(\tilde \kappa_i -s_i\,\tilde\lambda(\sigma_i)\big) \prod_{p=k+1}^n\bar\delta^{2|\cN} \big(\kappa_p-s_p\,\lambda(\sigma_p)\,|\,q_p-s_p\chi(\sigma_p)\big) \, .\nonumber
\end{equation}
This can be expressed in homogeneous coordinates on the Riemann sphere $\sigma_{\tilde\alpha}=\frac1s (1,\sigma)$ using the notation $(i \,j)=\sigma_{i\tilde\alpha}\sigma_j^{\tilde\alpha}$ (with indices raised and lowered by the usual skew symmetric $\epsilon_{\tilde\alpha\tilde\beta}$, but note that these here are not Lorentz spinor indices) as follows;
\be\label{dbar-sol-hgs}
Z(\sigma)= \sum_{i=1}^{k}\frac{\left(\kappa_{i},0,q_{i}\right)}{(\sigma\, \sigma_{i})}
\, , \qquad 
\tilde Z(\sigma)= \sum_{p=k+1}^{n}\frac{(\tilde{\kappa}_{p},0,0) }{(\sigma \,\sigma_{p})}\, ,
\ee
where we have rescaled $W$ and $Z$ by a factor of $1/s$.  Then 
\begin{multline}\label{final-form-hgs}
d\mu^{(\mathrm{4d})}_n= \frac{ \prod_{a=1}^n \rd^2 \sigma_a}{\vol \,\GL(2,\C)}   \,
\prod_{i=1}^k  \bar\delta^2\big(\tilde \kappa_i - \tilde\lambda(\sigma_i)\big)   \prod_{p=k+1}^n\bar\delta^{2|\cN} \big(\kappa_p-\lambda(\sigma_p)\,|\,q_p-\chi(\sigma_p)\big)\, .
\end{multline}
This defines a residue formula on $G(2,n)$, the $2n-4$-dimensional Grassmannian of $2$-planes in $n$-dimensions parametrized by the $\sigma_{a\tilde\alpha}$ modulo GL$(2,\C)$. This measure has several notable features:
\begin{itemize}
\item 
The $2n$ variables $\sigma_{a\tilde\alpha}$ are supported on the \emph{polarized scattering equations} 
\begin{equation}\label{PSC}
\tilde\lambda_{\dot\alpha}(\sigma_i)=\tilde \kappa_{i\dot\alpha}\, ,
\qquad \quad\lambda_\alpha(\sigma_p)
=\kappa_{p\alpha}\, ,
\end{equation}
for $i=1\ldots k$ and $p=k+1 \ldots n$.
These imply the ordinary scattering equations for $\sigma_a$; defining $P(\sigma)_{\alpha\dot\alpha}=\lambda(\sigma)_\alpha\tilde\lambda(\sigma)_{\dot\alpha}$ it is straightforward to show that $P$ has simple poles at $\sigma_a$ with residue $k_a$    and that $k_a\cdot P(\sigma_a)=0$ 
follows on the support of \eqref{PSC}. However, the polarized scattering equations also incorporate the choice of polarization data via the $s_a$, and are refined to give just those $A(n-3,k-2)$ solutions appropriate to N$^k$MHV degree, where $A(p,q)$ are the Eulerian numbers\footnote{Satisfying the recursion $A(p,q)=(p-q)A(p-1,q-1)+(q+1)A(p-1,q)$.}  \cite{Cachazo:2012da}.

\item For maximal supersymmetry, the measure $d\mu^{(\mathrm{4d})}_n$ only depends on the MHV sector, not the specific helicity assignment of the $n$ gluons. However, this `dihedral symmetry' of the N$^k$MHV amplitude is not manifest, and the above formul{\ae} require a choice of helicity assignment, here with the first $k$ particles of negative helicity.

\item The measure contains $2n$ bosonic delta functions but only $2n-4$ integrals; with the difference due to the $\mathrm{vol}\, (\GL(2,\C))$ quotient.  The remaining delta-functions encode momentum-conserving delta functions, as can be seen from
\begin{equation}
\sum_{p=k+1}^{n}\kappa_{p}\tilde{\kappa}_{p}=\sum_{p=k+1}^{n}\tilde{\kappa}_{p}\sum_{j=1}^{k}\frac{\kappa_{j}}{(p \, j)}=-\sum_{j=1}^{k}\kappa_{j}\tilde{\kappa}_{j}\,,
\end{equation}
where we used the first (second) set of delta functions in \eqref{final-form-hgs} to get the first (second) equality; similarly we obtain supermomentum conservation $\sum_{a=1}^{n}\tilde{\kappa}_{a}q_{a}=0$.

\item The amplitude formula \eqref{final-form} can be verified at $\cN=0$ by integrating out the moduli in equ.\ (3.22)  in \cite{Witten:2004cp}, 
see \cite{Geyer:2016nsh} for details.

\item While maximal supersymmetry with $\mathcal{N}=4$ naively seems to double the spectrum of the model, both vertex operators \eqref{YM'} are representatives of the \emph{same} space-time field, as established by the Penrose transform that maps, for example,   the same ASD Maxwell field to an element of  $H^1(\PT,\cO(-4))$ or to one in $H^1(\PT^*,\cO)$. Thus vertex operators $\cV'$ on $\P\T$ and $\tilde \cV$ on $\P\T^*$ represent the same space-time multiplet for $\cN=4$.

\item  The model also contains vertex operators for non-minimal conformal gravity states,  believed to coincide with the analogous states in the original twistor string \cite{Berkovits:2004hg}.  The full model is understood to give amplitudes for this combined $\cN=4$ super-conformal combination of conformal gravity and Yang-Mills theory, with the multi-trace terms in the current correlator corresponding to  interactions mediated by scalars of conformal supergravity.
\end{itemize}

\subsection{The twistor-string \texorpdfstring{formul{\ae}}{formul{\ae}}.} 
Soon after the original twistor-string, Roiban, Spradlin \& Volovich \cite{Roiban:2004yf} simplified its correlation function to give the (historically) first fully localized worldsheet formula for a field theory amplitude.
We present it out of historical order because it has additional moduli integrals compared to the formul{\ae} discussed so far.    This formula also requires $\cN=4$ supersymmetry for anomaly cancellation and for the expression to be well-defined. It does then have the benefit of manifesting the full dihedral symmetry of the N$^k$MHV amplitude irrespective of the specific helicities of individual gluons. In contrast, the ambitwistor string formul{\ae} do not manifest this symmetry.

The key distinction from the 4d ambitwistor-string is that the twistors $Z$ are understood to be sections of  line bundles $L\rightarrow \Sigma$ of degree $d >0$, and  the dual twistors $\tilde Z$ sections of $L^*\otimes K_\Sigma$. Moreover, only twistorial vertex operators are used so that it is more natural to refer to the model as a twistor-string.   At genus zero and  degree $d$, the maps $Z:\Sigma \rightarrow \P\T$ have moduli $M_r\in \T$, $r=0,\ldots ,d$ given by
\begin{equation}\label{moduli}
Z(\sigma)=\sum_{r=0}^d M_r\, \sigma^r\, , \qquad M_r=(m_{r\alpha},m_r^{\dot\alpha}, m^I_r)\, ,
\end{equation}
where $\sigma$ is an affine coordinate on $\Sigma=\CP^1$ and the $m^I_r$ are fermionic with the remaining components of the $M_r$ bosonic.

At $\cN=4$, the vertex operators \eqref{YM'} or \eqref{YM-vertex} contain the full multiplet including gluons of both helicities.  The full tree-level Yang-Mills super-amplitude is   obtained as the correlator of $n$ such vertex operators of one type or other.  There are no contractions between the vertex operators beyond the current algebra which give the usual Parke-Taylor factor, with the usual caveats about multi-trace terms.  Thus the path-integral immediately  localizes onto the zero-modes \eqref{moduli}, yielding the amplitude formula
\begin{equation}\label{moduli-formula}
\cA_n=\int \frac{d^{4d+4|4d+4} M\, }{\vol \mathrm{GL}(2)}
 \prod_{a=1}^n \frac{d\sigma_a\, ds_a }{s_a}\;\bar\delta^2\big(\kappa_a -s_a\lambda(\sigma_a)\big) \, \e^{is_a([\mu\tilde\kappa_a] +\chi^I \tilde q_{aI})} \; \Pt(\alpha) \, .
\end{equation}
In order to simplify this further, we can formally integrate out the moduli $m_r^{\dot\alpha}, m_r^I$. Since they only appear in the exponentials, this leads to additional delta-functions,
\begin{equation}\label{RSVW}
\cA_n=\int  \frac{d^{2d+2}m_{r\alpha}\, }{\vol \mathrm{GL}(2)} 
\prod_{a=1}^n \frac{d\sigma_a\, ds_a }{s_a} \; \bar\delta^2\big(\kappa_a -s_a\lambda(\sigma_a)\big) \,    \prod_{r=0}^d \delta^{2|4}\Big(\sum_{a=1}^n \sigma^r_a s_a\tilde\kappa_{a\alpha}\Big) \;\Pt(\alpha)\, .
\end{equation}
In this formula,  \eqref{moduli} gives $\lambda(\sigma_a)=\sum_r m_{r\alpha}\sigma_a^r$.
Once the GL$(2)$ quotient is taken into account,  there are four more delta-functions than integrals encoding supermomentum conservation. This can be made  explicit by summing the arguments of the $(d+1)$  $\bar \delta^{2|4}$-functions, multiplied by $m_{r\alpha}$, and using the support of the first $n$ delta-functions.
Thus the  moduli integrals over $(s_a, \sigma_a, m_{r\alpha})$ modulo GL$(2,\C)$  can be performed against  the remaining delta functions to give a sum of residues  multiplied by the momentum conservation delta-functions.  This is analogous to the 4d-ambitwistor string, but with an additional $2d+2$ moduli integrals and delta-functions.

\begin{itemize}
\item The fermionic delta functions relate the MHV degree $k$ of the amplitude to the degree of the line bundle  by $k=d-1$

\item In \cite{Roiban:2004yf, Witten:2004cp}, the formula was shown to be parity invariant, which is far from manifest, with  a number of further checks.  A BCFW recursion proof was given in \cite{Skinner:2010cz}. 

\item In \cite{Berkovits:2004tx} it was shown that one can further introduce vertex operators for $\cN=4$ superconformal gravity.  There are two multiplets, one containing the ASD Weyl tensor, determined by a divergence-free vector field $f(Z)\cdot\p_Z$ and realized in the twistor-string by the vertex operator $f\cdot \tilde Z$, and one containing the SD Weyl tensor, given by a 1-form $g(Z)\cdot \p Z$. Correlators involving these states are more complicate because  the $\tilde Z$ factors must be contracted completely before one can reduce to a moduli integral.\footnote{Only   $f\cdot \tilde Z$ and its conjugate need to be used in the 4d ambitwistor model.}

\item
The equivalence with the 4d ambitwistor formula was proved in \cite{Geyer:2016nsh}; it essentially uses the moduli integrals in  the twistor string formula  \eqref{moduli-formula} to perform a \emph{twistor-transform} of $d+1$ of the  cohomology classes in the vertex operators.  This for example maps an element of  $H^1(\PT,\cO(-4))$ describing an ASD Maxwell field to that in $H^1(\PT^*,\cO)$  for the same field. 

\item Both in the twistor string and in the ambitwistor string, a number of open questions remain. In the twistor string, we do not  include vertex operators $\tilde{\mathcal{V}}_a$ constructed from dual twistor cohomology classes, but there does not seem to be  a good reason not to.   Related to this, gauging the current $Z\cdot \tilde{Z}$ in the 4d ambitwistor string should also result in a sum over the degree of line bundle associated to the gauge field $a$, but this doesn't seem to play a role in the amplitude formul{\ae}; see also 
\cite{Geyer:2016nsh}.
\end{itemize}

\subsection{Einstein supergravity models and amplitudes.}  The twistor string for $\cN=8$ Einstein supergravity was introduced by Skinner \cite{Skinner:2013xp} as the model underpinning the earlier  Cachazo-Skinner formula \cite{Cachazo:2012kg} (proved in \cite{Cachazo:2012pz}).  A 4d ambitwistor version was then introduced in \cite{Geyer:2014fka}.  Again, both have essentially the same underlying worldsheet model but with different worldsheet assignments of twists  for the twistor and dual twistor target fields and  ghosts.  We focus here on the 4d ambitwistor version for brevity;   see  \cite{Cachazo:2012kg, Skinner:2013xp} for the twistor-string version.  These have the advantage of full permutation invariance, at the price of additional moduli integrals.

 In order to break conformal invariance we introduce skew bilinear forms on twistor space and its dual, the
infinity twistor and its dual. 
When non-degenerate, these  encode a cosmological constant and a gauging of $R$-symmetry. Although the model was originally introduced  incorporating a cosmological constant, here we restrict to the Minkowski space model and its amplitudes.
In this case the infinity twistor  have rank 2  and we will denote contractions with a pair of twistors by $\la Z_1, Z_2\ra:=\la \lambda_1\, \lambda _2\ra$ and with a pair of dual twistors by $[\tilde Z_1,\tilde Z_2]:=[\tilde\lambda_1\, \tilde\lambda_2]$.

We introduce a worldsheet superalgebra by extending the target to include $(\rho,\tilde \rho)\in\T\times \T^*$ that are parity reversed, taking values in $\C^{\cN| 4}\otimes K^{\nicefrac{1}{2}}_\Sigma$ (rather than $\C^{4|\cN}$); they are taken to be worldsheet spinors  in both the  twistor-string and the 4d ambitwistor string versions.  An additional set of  constraints are gauged, and including the original $Z\cdot\tilde Z$, the total set of currents that are gauged in this model becomes  
\begin{equation}
K_{\mathfrak a}=\left(Z\cdot \tilde Z, \rho\cdot \tilde \rho,  \tilde Z\cdot \rho, [\tilde Z, \tilde \rho], Z\cdot \tilde \rho, \la Z, \rho\ra, \la \rho, \rho\ra, [\tilde \rho, \tilde \rho]\right)\,.
\end{equation}
In the BRST quantization, this introduces corresponding  ghosts $(\beta_{\mathfrak a},\gamma^{\mathfrak a})$, 
together with the fermionic $(b,c)$ ghosts as before \cite{Adamo:2013tca}, and leads to a BRST $Q$-operator
\begin{equation}
Q=\int c T + \gamma^{\mathfrak a}K_{\mathfrak a} -\frac i2 \beta_{\mathfrak a} \gamma^{\mathfrak b}\gamma^{\mathfrak c}\,  C^{\mathfrak a}_{\mathfrak bc}\, ,
\end{equation}
where $C^{\mathfrak a}_{\mathfrak bc}$ are the structure constants of the current algebra $K_{\mathfrak a}$.  The model again has a potential gauge anomaly, whose coefficient vanishes for $\cN=8$ maximal supersymmetry.

In these Einstein gravity models,   $Q$-invariance implies that vertex operators are built from $\dbar$-closed $(0,1)$-forms $h$ of weight two on twistor space, as well as $\tilde h$ on dual twistor space for the ambitwistor string.  
For momentum eigenstates, $h$ and $\tilde{h}$ are
$$
 h_a=\int \frac{\rd s_a}{s_a^3}\bar\delta^{2|\cN}(\kappa_a-
 s_a\lambda|q_a-s_a\chi)\,\e^{is_a[\mu\,
  \tilde\kappa_a]}\, , \quad 
 \tilde{h}_a=\int\frac{\rd s_{a}}{s_{a}^{3}}\bar{\delta}^{2}(\tilde{\kappa}_{a}-s_{a}\tilde{\lambda})\, \e^{is_{a}\left(\la\tilde{\mu}\,\kappa_{a}\ra+\tilde{\chi}_{r}q_{a}^{r}\right)}\,.
$$
These yield two types of vertex operators, appearing in integrated or fixed form, here integration being with respect to ghost zero modes.   The ghosts $\gamma=(\gamma^3,\gamma^4)$, $\nu=(\gamma^5,\gamma^6)$ each have one zero mode that are fixed by the insertion of one each of
\begin{equation}
 V_h=\int_{\Sigma}\delta^2(\gamma)h\, , \qquad  \widetilde V_{\tilde{h}}=\int_{\Sigma}\delta^2(\nu)\tilde{h}\,.
\end{equation}
The remaining particles are represented by integrated vertex operators
\begin{equation}\label{eq:V_int_4d}
 \mathcal{V}_h=\int \Big[\tilde Z,\frac{\p h}{\p Z}\Big]+\Big[\tilde{\rho},\frac{\p}{\p Z}\Big]\,\rho\cdot\frac{\p h}{\p Z},\qquad
 \widetilde{\mathcal{V}}_{\tilde{h}} =\int\Big\la Z,\frac{\p \tilde{h}}{\p \tilde Z}\Big\ra+\Big\la\rho,\frac{\p}{\p \tilde Z}\Big\ra\tilde{\rho}\cdot\frac{\p \tilde{h}}{\p \tilde Z}.
\end{equation}
Amplitudes are now given by the worldsheet correlation function
\begin{equation}
 \cM=\left\langle \widetilde V_{\tilde h_1}\prod_{i=2}^{k} \widetilde{\mathcal{V}}_{\tilde h_i} \prod_{p=k+1}^{n-1} \mathcal{V}_{{h}_p} V_{{h}_n}\right\rangle.
\end{equation}
The correlator of  the $(\rho,\tilde \rho)$ fermion system  is given by the determinant of the following $n\times n$ matrix:
\begin{equation}
\cH= \begin{pmatrix}{ \mathbb{H}}& 0\\ 0&\widetilde{\mathbb{H}}\end{pmatrix}\,,
\end{equation}
where, for $i,j\in\{1,...,k\}$ and $p,q\in\{k+1,...,n\}$
\begin{align}
 {\mathbb{H}}_{ij}=\begin{cases}\frac{\braket{i\, j}}{(i\, j)}, &  i\neq j, \\ - \sum_{l=1,l \neq i}^{k} {\mathbb{H}}_{il}\, , &i=j\, , \end{cases}
  \qquad
 \widetilde{\mathbb{H}}_{pq}=\begin{cases}\frac{[p\, q]}{(p\, q)}, & p\neq q, \\ -\sum_{r=k+1,r \neq p}^n \widetilde{\mathbb{H}}_{pr}\, , & p=q\, . \end{cases}
\end{align}
The off-diagonal element $\cH_{ij}$ comes from the contraction of the $\rho$-term in the $i$th vertex operator with the $\tilde \rho$-term in the $j$th, and the diagonal elements of $\cH$ stem from the first term in the integrated vertex operator \eqref{eq:V_int_4d}. By an analogous calculation as for Yang-Mills theory, we then obtain the  gravity amplitudes 
\begin{equation} \label{Mgrav}
 \cM_n=\int d\mu^{(\mathrm{4d})}_n \;\text{det}'(\cH)\,,
\end{equation}
where $\det' \cH$ is the determinant omitting a row and column from each of $\widetilde{\mathbb H} $ and $\mathbb{H}$ corresponding to the fixed vertex operators;  the answer is independent of this choice because each has kernel $(1,\ldots,1)$.

\begin{itemize}
\item 
Through lack of space we do not include the Cachazo-Skinner twistor-string gravity formula, but refer the reader to \cite{ Cachazo:2012kg,Cachazo:2012pz, Skinner:2013xp}.
The equivalence between  these formul{\ae} above follows the corresponding story for Yang-Mills, see  \cite{Geyer:2016nsh}.

\item It is possible to consider higher rank versions of the infinity twistor with bosonic rank four corresponding to the inclusion of a cosmological constant \cite{Penrose:1986ca} and higher rank in the fermionic directions corresponding to gauged supergravity theories with gauged $R$-symmetry \cite{Wolf:2007tx}. The Skinner model \cite{Skinner:2013xp} was originally formulated in this way. Corresponding amplitude formul{\ae} with non-zero cosmological constant have been explored in \cite{Adamo:2013tja,Adamo:2015ina, Adamo:2021bej} but remain conjectural.

\item Comparison between the Yang-Mills versus gravity formula and the corresponding  CHY formula restricted to 4d makes clear that the CHY Pfaffian must be equal to $\det{}'\cH$.  This is shown via analyticity and CFT arguments  in \cite{Roehrig:2017wvh} and used to extend the formul{\ae} to include Einstein Yang-Mills amplitudes improving \cite{Adamo:2015gia}.

\item The double copy is \emph{not} manifest in these formul{\ae}. To see it, one must integrate out the $s_a$ coordinates in the polarized scattering equations measure.  This yields a second $\det{}'\cH$ multiplied by the CHY measure, providing a geometric origin to this second `copy' in the gravity formul{\ae} \cite{Albonico:2020mge}. See also \cite{He:2016dol,Zhang:2016rzb} which include some interesting extensions.

\item The polarized  scattering equations have been  extended to higher dimensions, including 6d in \cite{  Geyer:2018xgb, Albonico:2020mge}, 10 \& 11d in \cite{Geyer:2019ayz}, with underlying models in 5 \& 6d in \cite{Geyer:2021oox}. 
These 6d formul{\ae} are distinct from the earlier ones of \cite{Heydeman:2017yww, Heydeman:2018dje, Cachazo:2018hqa} which give twistor-string like expressions for D5 and M5-branes.  The corresponding Yang-Mills and gravity formul{\ae} however become awkward for odd numbers of particles, an issue that  doesn't arise in the brane theories, since brane amplitudes are only nontrivial for even numbers of particles.

\item Twistorial models in 10d  that solve the $P^2=0$ constraint have been introduced using `impure' 10d  twistors in \cite{ Berkovits:2019bbx, Reid-Edwards:2017goq} and pairs of pure twistors in \cite{Sepulveda:2020wwq}.

\item There exists a two-twistor representation of the space of massive particles \cite{Penrose:1974di} and its complexification can be used as a target space for a massive twistor string \cite{Albonico:2022pmd}.  This yields massive amplitudes in four dimensions  including  half-integral spins and manifest supersymmetry in formul{\ae} localized again on the polarized scattering equations.  These formul{\ae} also arise by dimensional reduction from 6d and were first found in \cite{ Albonico:2020mge} (again preceded by related massive amplitude formul{\ae} in \cite{Cachazo:2018hqa}). Such massive formul{\ae} have also been used to construct 1-loop integrands \cite{Wen:2020qrj}.

\item
Both the twistor-string and the ambitwistor-string formul{\ae} naturally embed \cite{Bullimore:2009cb,Arkani-Hamed:2009kmp} into the Grassmannians of \cite{ArkaniHamed:2009dn}, where they can be related to twistor-string formul{\ae} for leading singularities and BCFW terms. In 6d a Lagrangian Grassmannian approach LG$(n,2n)$ was proposed in \cite{Cachazo:2018hqa} that facilitates the comparison \cite{Schwarz:2019aat} between the different 6d amplitude formul{\ae} \cite{Cachazo:2018hqa} and \cite{Geyer:2018xgb}.

\end{itemize}


\section{Loop amplitudes at higher genus}\label{Loops}
The ambitwistor string not only provides a beautiful geometric explanation of the CHY formul{\ae} but it   allows  us to extend these tree-level amplitude formul{\ae} in a variety of directions. One important such avenue are extensions of the worldsheet CHY formul{\ae} to loop amplitudes.
 On a practical level, it is clear that the only way to approach this problem is via a \emph{model} --- guessing  loop-level formul{\ae} is simply not generally feasible.  

\begin{figure}[ht]
\begin{center}
 \includegraphics[width=12cm]{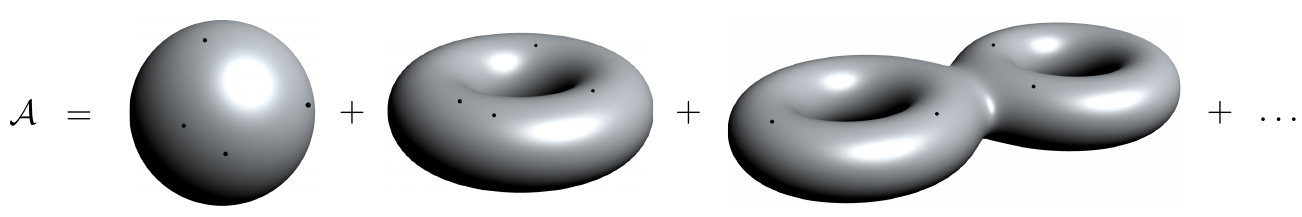}
 \caption{Schematic loop expansion of the amplitude. In worldsheet models such as the ambitwistor string, $g$-loop amplitudes correspond to correlators on genus-$g$ Riemann surfaces.}
 \label{fig:amplitude_WS}
\end{center}
\end{figure}

In the first instance, as a closed-string worldsheet model, we will see that  $g$-loop amplitudes  in the ambitwistor string should be  given by  correlators over genus-$g$ Riemann surfaces; this is already a significant simplification from the \emph{super} Riemann surfaces usually encountered in the superstring. The full amplitude then has a perturbative expansion  as a sum over topologies, schematically expressed in \cref{fig:amplitude_WS}.

In this section, we aim to give a sketch of the correspondence between higher-genus correlators and loop amplitudes, many additional details can be found in the original papers carrying out the calculations at one 
\cite{Adamo:2013tsa,Casali:2014hfa}
and two loops \cite{Adamo:2015hoa, Geyer:2018xwu}. 
The focus, both here and in the original papers, lies on the RNS model $S_{\mathrm{II}}$ describing  supergravity, because the others models are `contaminated' by their unphysical gravitational states propagating in the loop.\footnote{Progress for the twistorial models has also been rather limited, see  \cite{Dolan:2007vv, Farrow:2017eol, Wen:2020qrj}.} Rather than providing a fully self-contained derivation of the loop amplitudes, we emphasize here the key features distinguishing the ambitwistor string from the superstring, and reflecting its \emph{field theory} nature. We will encounter this repeatedly via calculations that bear, superficially, a close similarity to the superstring \cite{DHoker:2001kkt, DHoker:2001qqx, DHoker:2001foj, DHoker:2001jaf, DHoker:2005dys, DHoker:2005vch, DHoker:2002hof}, but with considerable simplifications and important differences, all originating in the chiral nature of the ambitwistor string and its supersymmetry structure --- the worldsheet super gauge algebra discussed in \cref{sec:models},  as opposed to the worldsheet super-diffeomeorphisms familiar from the superstring.  In  the superstring,  the fermionic constraint squares to give the stress-energy tensor generating worldsheet diffeomorphisms, so the correlator is an integral over the moduli space of a supermanifold. In the ambitwistor string on the other hand, the worldsheet fermionic symmetry squares to give $P^2$ which, in generating translations along the geodesics, does not complicate  the moduli space of the Riemann surface.

\subsection{From higher-genus correlators to loop amplitudes}
The ambitwistor string provides a clear prescription for calculating loop amplitudes as a correlator over a higher-genus worldsheet. Our first goal will be to understand this prescription, before we proceed to evaluate the correlator.

\begin{figure}[ht]
\begin{center}
 \includegraphics[width=6cm]{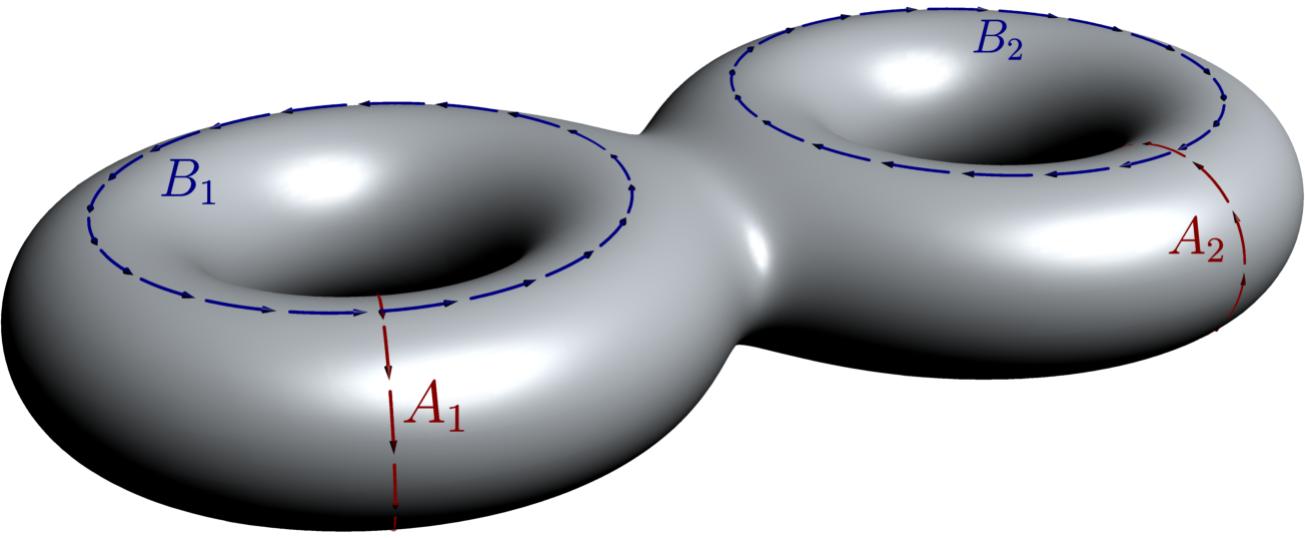}
 \caption{The homology basis at genus two.}
 \label{fig:genus2}
\end{center}
\end{figure}

\paragraph{The correlator:} The genus-$g$ ambitwistor string  correlator with $n$ vertex operators involves an integral over the moduli space $\mathfrak{M}_{g,n}$, stemming from integral over inequivalent metrics under conformal transformations \cite{DHoker:1988ta, Polchinski:1998rq}.  A convenient description of this moduli space (up to genus three) is given by the \emph{period matrix}, which is defined as follows. For a genus-$g$ Riemann surface, we choose a homology basis of cycles $A_I$ and $B_I$, with $I = 1, . . . , g$ such
that the intersection form is canonical, 
as illustrated in \cref{fig:genus2} for genus two.
Transformations acting on the homology basis $(A_I, B_I)$ while leaving the intersection form invariant are known as modular transformations, and form the \emph{modular group} Sp$(2g,\mathbb{Z})$.
If we normalize the holomorphic abelian differentials $\omega_I$ against the $A$-cycles, then the period matrix $\Omega_{IJ}=\Omega_{(IJ)}$ is given by the $B$-periods;
\begin{equation}
 \oint_{A_I}\omega_J = \delta_{IJ}\,,\qquad\qquad \oint_{B_I}\omega_J = \Omega_{IJ}\,.
\end{equation}
Up to genus three, $\mathfrak{M}_{g,n}$ can then be parametrized by such period matrices $\Omega$ up to the modular group as the dimension count  for these two spaces agrees up to genus three, with $\mathrm{dim}\,\mathfrak{M}_{g,n} = 3g-3$ and $ \mathrm{dim} \,\Omega = \frac{1}{2}g(g+1)$.
This is particularly simple in the case of the torus, where the moduli space is parametrized by the complex parameter $\tau \in \mathcal{F}$, where   
\begin{equation}
\mathcal{F}=\{\tau\in\mathbb{C}|\,|\tau|\geq1, \,-\frac{1}{2}\leq \mathfrak{Re}(\tau)\leq\frac{1}{2}\}\,,
\end{equation}
is  the fundamental domain  obtained by quotienting the upper half-plane by the Dehn twists generating the modular group PSL$(2,\mathbb{Z})$, see  \cref{fig:residue}.

At loop level, the GSO projection --  required to restrict to the correct supergravity degrees of freedom -- results in a sum over worldsheet spin structures $\kappa\in (\mathbb{Z}/2\mathbb{Z})^{2g}$. 
Since the GSO projection is implemented independently for each of the two chiral spinors, this leads to a double sum over spin structures whose relative phases $\eta_\kappa$,  $\eta_{\tilde \kappa}$ are entirely determined by modular invariance and unitarity \cite{Seiberg:1986by}. 
The $n$-particle $g$-loop ambitwistor string correlator can then be schematically
expressed for $g>1$ as follows:
\begin{equation}
 \mathcal{A}_{n}^{(g)} = \int_{\mathfrak{M}_{g,n}}\prod_{I\leq J}d\Omega_{IJ}\sum_{\kappa,\tilde\kappa} \eta_\kappa\eta_{\tilde\kappa}
 \left\langle  
 \prod_{r=1}^{3g-3}  b_r\tilde b_r
  \,\bar\delta( \langle \mu_r P^2\rangle\,)
 \prod_{\alpha=1}^{2g-2} \Upsilon_\alpha\tilde\Upsilon_\alpha
 \prod_{i=1}^n \mathcal{V}_i \right\rangle_{\kappa\tilde\kappa}\,.
\end{equation}
The torus is exceptional needing  one fixed vertex operator due to the remaining symmetry and associated zero mode of the $c$ and $\tilde c$ ghosts; the first product are then nontrivial with a single entry.
Here  $\mu_r\in \Omega^{(0,1)}\otimes T^{(1,0)}_\Sigma$ is a  Beltrami differential, and $\Upsilon_\alpha$, $\tilde{\Upsilon}_\alpha$ are the \emph{picture changing operators} introduced in \cref{sec:VO}.

A key feature is that the additional delta-functions, $\bar\delta(\langle\mu_r P^2\rangle)$  enforces that $P^2$ vanishes. These form an important part of the loop-level scattering equations, as we will see below, and give $n_b$ new constraints, with  $n_b=1$ on the torus and $n_b=3g-3$ at higher genus leading again to a sum over residues as at tree-level. Moreover, we note that for $g\geq 1$,  not all picture-changing operators $\Upsilon_\alpha\tilde\Upsilon_\alpha$ can be absorbed by the vertex operators; this will affect the structure of the integrand.

\paragraph{Scattering equations:} Since the correlator only depends on $X$ via the exponentials in the vertex operators, the $PX$-path integral can be evaluated following the same strategy as at tree-level. As discussed in \cref{sec:models}, 
the $X$ zero-mode integral then gives $d=10$ momentum-conserving delta-functions, and the non-zero modes localize $P$ onto the solution to its equations of motion \eqref{eq:EoM_P}. Although the genus-$g$ Green's function for the $PX$-system is complicated,\footnote{Appendix A of ref.~\cite{DHoker:2001qqx} serves as an excellent summary for CFTs on higher-genus worldsheets.} the equations of motion fortunately have a simple solution in terms of holomorphic and meromorphic differentials;
\begin{equation}\label{eq:P_g}
 P_\mu(z) = \ell^I_\mu\, \omega_I + \sum_i k_{i\mu}\,\omega_{i,*}(z)\,.
\end{equation}
It is traditional at higher genus to use the worldsheet coordinate $z$ rather than $\sigma$ to avoid confusion when simplifying the resulting amplitude formul{\ae}, see \cref{nodal}.  Here, in the second term, the meromorphic differentials $\omega_{ij}$ are so-called  \emph{ Abelian differentials of the third kind}, i.e.,  1-form in $z$, with a simple poles at two marked points $z_i$, $z_j$ with respective residues $\pm 1$;  $z_*$ is a reference point, but momentum conservation ensures that $P$ has no pole at that point. 
The first term arises because,  at loop level, the equations $\dbar P=0$ admits homogeneous solutions $\ell^I_\mu \omega_I$,  given by the holomorphic differentials $\omega_I$ multiplied by the zero-modes $\ell^I_\mu$. The path integral now includes an integral over these zero-modes, named  to suggest already their role as the loop momentum of the amplitude.

The scattering equations -- both geometrically and from the CFT point-of-view -- still play the same role as at tree-level, enforcing the constraint $P^2=0$, now however on the genus-$g$ worldsheet.  As evident from the correlator, there are now two types of constraints: $n$ scattering equations (or $n-1$ on the torus, due to its translation invariance) are still associated to the moduli of the insertion points of the vertex operators, and resemble the tree-level constraints,
\begin{equation}
 \mathcal{E}_i = \mathrm{Res}_i\, P^2 = 2k_i\cdot P(z_i)\,.
\end{equation}
In contrast to tree-level however, these constraints are not sufficient to ensure $P^2=0$ on higher-genus surfaces due to the homogeneous solutions.
This is reflected in the additional $n_b$ delta-functions in the correlator, which provide the remaining constraints. They are best parametrized by expanding $P$, on the support of the remaining scattering equations $\mathcal{E}_i$, into a basis of quadratic holomorphic differentials,
\begin{equation}
 P^2 = u \, dz^2\,,\qquad P^2 = u^{IJ} \, \omega_I\omega_J\,,
\end{equation}
given here at one and two loops respectively. The new, loop-level scattering equations then enforce $u=0$ and $u_{IJ}=0$ respectively. For fixed loop momenta $\ell^I$, these constraints are solved by localizing the moduli $\Omega_{IJ}$ themselves so that, as we will see, the \emph{loop integrand} reduces to a sum of residues in $\mathfrak{M}_{g,n}$.

\paragraph{Loop amplitudes:} The ambitwistor string thus leads to the following formul{\ae},
\begin{equation}\label{eq:A_g=1}
 \cA_{n}^{(g)} = \delta^{10}\Big(\sum_{i=1}^n k_i\Big)\; 
 \int \prod_{I=1}^gd^{10}\ell_I \; \; \mathfrak{I}_{n}^{(g)}\,,
\end{equation}
where the integral is over the zero-modes $\ell^I_\mu $ of $P_\mu$, and the `loop integrand' $\mathfrak{I}_{n}^{(g)}$ is given by an integral over the moduli space $\mathfrak{M}_{g,n}$, \footnote{Note that for the torus, this only contains $n-1$ scattering equations $\mathcal{E}_i=\mathrm{Res}_iP^2$ as discussed above.}
\begin{equation}\label{eq:integrand_g}
 \mathfrak{I}_{n}^{(g)} =  \int_{\mathfrak{M}_{g,n}} \prod_{I\leq J}d\Omega_{IJ}\; \bar\delta(u_{IJ})\,\prod_{i}\bar\delta\left(\mathrm{Res}_i P^2\right)\; \mathcal{I}_n^{\scalebox{0.6}{chi}}\, \tilde{\mathcal{I}}_n^{\scalebox{0.6}{chi}}\,.
\end{equation}
The chiral integrands $\mathcal{I}_n^{\scalebox{0.6}{chi}}$ stem from the remainder of the correlator, including the sum over spin structures, and contain the  non-trivial chiral partition functions $ \mathcal{Z}^{\scalebox{0.6}{chi}}_\kappa$ of all fields, as well as Pfaffian factors from the $\Psi$, $\tilde\Psi$ correlators.

For brevity, we only give explicit expressions for even spin structures at one loop; this covers many cases of interest, as odd spin structures don't contribute in $d\leq 9$ dimensions or for four particles at $g\leq 3$. Formul{\ae} for odd spin structures and two-loops can be found in \cite{Adamo:2013tsa} and \cite{Adamo:2015hoa, Geyer:2018xwu} respectively. For an even spin structure $\delta$, 
\begin{equation}\label{eq:int_1loop}
 \mathcal{I}_n^{\scalebox{0.6}{chi}} = \sum_\delta (-1)^{\delta}\,\mathcal{Z}^{\scalebox{0.6}{chi}}_\delta\; \Pf \big(M_\delta\big)\,,\qquad \mathrm{with}\;\;
 M_\delta=\begin{pmatrix} \,A  &-C^T\\ \,C    &\;\;\,B
 \end{pmatrix}\,,
\end{equation}
where the chiral partition function $ \mathcal{Z}^{\scalebox{0.6}{chi}}_\delta={\vartheta_\delta(0|\tau)^4}\,  {\eta(\tau)^{-12}}$ is given in terms of Jacobi theta-functions  and the Dedekind eta-function, and where $M_\delta$ is a $2n\times 2n$ matrix  with components $A_{ii}=B_{ii}=0$, $C_{ii} = -\epsilon_i\cdot P(z_i)$ and
\begin{equation}\label{eq:M_1loop}
 A_{ij}=k_i\cdot k_j \, S_\delta(z_{ij}|\tau)\,,\quad
 C_{ij}=\epsilon_i\cdot k_j \, S_\delta(z_{ij}|\tau)\,,\quad
 B_{ij}=\epsilon_i\cdot \epsilon_j \, S_\delta(z_{ij}|\tau)\,.
\end{equation}
The structure of the matrix is reminiscent of that at tree-level, but its components now depend on the punctures $z_i$ via the  Green's function for the fermion $\Psi$-system; the Szeg\H{o} kernel
\begin{equation}
 S_\delta(z-w|\tau) = \frac{\vartheta_1(0|\tau)}{\vartheta_1(z-w|\tau)}\frac{\vartheta_\delta(0|\tau)}{\vartheta_\delta(z-w|\tau)}\sqrt{dz}\sqrt{dw}\,.
\end{equation}

\paragraph{Four particle amplitude: } For four external particles, these amplitude expressions simplify drastically, and the sum over spin structures can be performed explicitly. In this case the chiral integrand is commonly written as $ \mathcal{I}_4^{\scalebox{0.6}{chi}}=\mathcal{K}\, \mathcal{Y}^{\scalebox{0.6}{$(g)$}}$, where $\mathcal{K}$ is a universal kinematic prefactor defined by the tree amplitudes $\mathcal{A}_4^{(0)}= \mathcal{K}\tilde{\mathcal{K}} / stu$, and
\begin{equation}
 \mathcal{Y}^{\scalebox{0.6}{$(1)$}} =\mathcal{K} \prod_{i=2}^4 dz_i\,,\qquad
 \mathcal{Y}^{\scalebox{0.6}{$(2)$}} =\mathcal{K} \left(s\, \Delta_{14}\Delta_{23} - t\,\Delta_{12}\Delta_{34}\right)\,.
\end{equation}
Here, $\Delta_{ij}=\varepsilon^{IJ}\omega_I(z_i)\omega_J(z_j)$ denotes the natural modular form of weight $-1$ on a $g=2$ Riemann surface, and $s=2 k_1\cdot k_2$ and $t=2k_1\cdot k_4$ are Mandelstam variables.

\subsection{Properties and further developments}
Since supergravity amplitudes are -- in contrast to the superstring -- not UV finite, the ambitwistor string correlators \eqref{eq:A_g=1} are divergent and so should be understood as formal expressions only before regularization. The natural object of interest is the \emph{loop integrand} $\mathfrak{I}_{n}^{(g)}$ and is well-defined. 
This loop integrand has indeed many interesting properties:

\begin{itemize}
 \item First and foremost, $\mathfrak{I}_{n}^{(g)}$ is an integral over the moduli space $\mathfrak{M}_{g,n}$ of marked Riemann surfaces. Aspects of Modular invariance, while far from manifest, follow as a consequence of the GSO projection and the resulting sum over spin structures \cite{Adamo:2013tsa, Geyer:2018xwu}. However, modular transformations affect the loop momentum $\ell^I$, so \eqref{eq:integrand_g} should be interpreted in the first instance as an integral over a specific fundamental domain and the integrand is not in itself expected to be modular invariant; full modular invariance is only restored after accounting for the transformation properties of the loop momenta. 
 Unlike in the superstring case,  modular invariance does \emph{not} guarantee finiteness of the amplitude, and, as we will see in the next section, we are dealing with field theory loop integrands, so that the moduli integral over the zero modes $\ell^I$ that remain to be performed are infinite.
 
 \item 
In analogy with the tree-level ambitwistor string correlator, and in contrast with the superstring, the moduli space integrals in $\mathfrak{I}_{n}^{(g)}$ are \emph{fully localized} on solutions to loop-level scattering equations. This is evident in the explicit formul{\ae} \eqref{eq:integrand_g}, but also has a clear CFT origin since both the integration measure and the delta-function insertions are  tied to the dimension of the moduli space $\mathrm{dim}\, \mathfrak{M}_{g,n} = n+3g-3$.
 \item 
The correlator  formul{\ae} \eqref{eq:integrand_g} also exhibit a  tantalizing resemblance to the superstring amplitude in the chiral splitting formalism \cite{DHoker:1988ta, DHoker:1989cxq}. This manifests itself both via closely analogous calculations (compare, for example, the two-loop calculations in \cite{DHoker:2001kkt, DHoker:2001qqx, DHoker:2001foj, DHoker:2001jaf, DHoker:2005dys, DHoker:2005vch} and \cite{Geyer:2018xwu}) and in the final answer; in fact the \emph{same} chiral half-integrands ${\mathcal Y}^{(g)}$ appear in the superstring 
\begin{equation}
 \label{eq:ssamp}
 {\mathcal A}_{\,\mathbb{S}}^{(g)} =  \mathcal{K}\tilde{\mathcal{K}}\int_{{\mathfrak M}_{g,4}} \Big|\! \prod_{I\leq J} d\Omega_{IJ\,} \! \Big|^2 \int \!\! d\ell \,\, \big|{\mathcal Y}^{(g)}\big|^2 \;\prod_{i<j} | E(z_i,z_j) |^{\frac{\alpha'  \!s_{ij}}{2}}\; 
|C(\ell^I)|^2\,,
\end{equation}
and the ambitwistor string for $n=4$ particles. Here the product over the prime form $E(z_i,z_j)$ and the chiral splitting factor $C(\ell^I)$ form the loop-level Koba-Nielsen factor characteristic of  the string. Gaining a better understanding of the close similarity between such different theories -- the superstring and supergravity -- still remains an open problem.  Various recent developments have aimed to clarify various aspects, such as the role of loop momentum in string theory  \cite{Tourkine:2019ukp}, the ambitwistor string moduli integrals \cite{Ohmori:2015sha}, detail of the chiral splitting in the ambitwistor string \cite{Kalyanapuram:2021vjt}, and the relation to other chiral strings \cite{Siegel:2015axg, Casali:2016atr, Casali:2017zkz, Yu:2017bpw}. 

\end{itemize}

The obvious counterpart to the problem of relating the ambitwistor string and the superstring concerns the field theory status of these formul{\ae}. 
Many important properties of field theory integrands are obscured on the higher genus worldsheet; in particular, it remains mysterious how expressions such as \eqref{eq:integrand_g} could possible give rise to the  simple rational functions of the external kinematics and loop momenta that constitute the supergravity integrand. In the words of the original paper \cite{Adamo:2013tsa}, this would surely require `miraculous simplifications'. In the next section, we will see investigate these `miraculous simplifications', and explore many of their consequences.


\section{From higher genus to the nodal sphere}
\label{nodal}
Despite the compactness of the formul{\ae}, the higher genus correlators are hard to compute and would seem to give highly transcendental functions rather than the rational functions we expect for field theory intgrands.  Remarkably one can  relate the complicated higher-genus correlators of the previous section to the familiar rational functions of field theory integrands. In doing so we obtain a new formulation for loop integrands on the Riemann sphere but with an extra node or double point for each loop order.  We also gain control over which fields run in the loop and can therefore generate loop integrand formul{\ae} for the full range  of theories in different dimensions that the CHY framework applies to, taking us far beyond the  restriction to type II supergravity in 10d of the previous section. 

\subsection{Residue theorem to the nodal sphere}
\begin{figure}[ht]
\begin{center}
  \includegraphics[width=14cm]{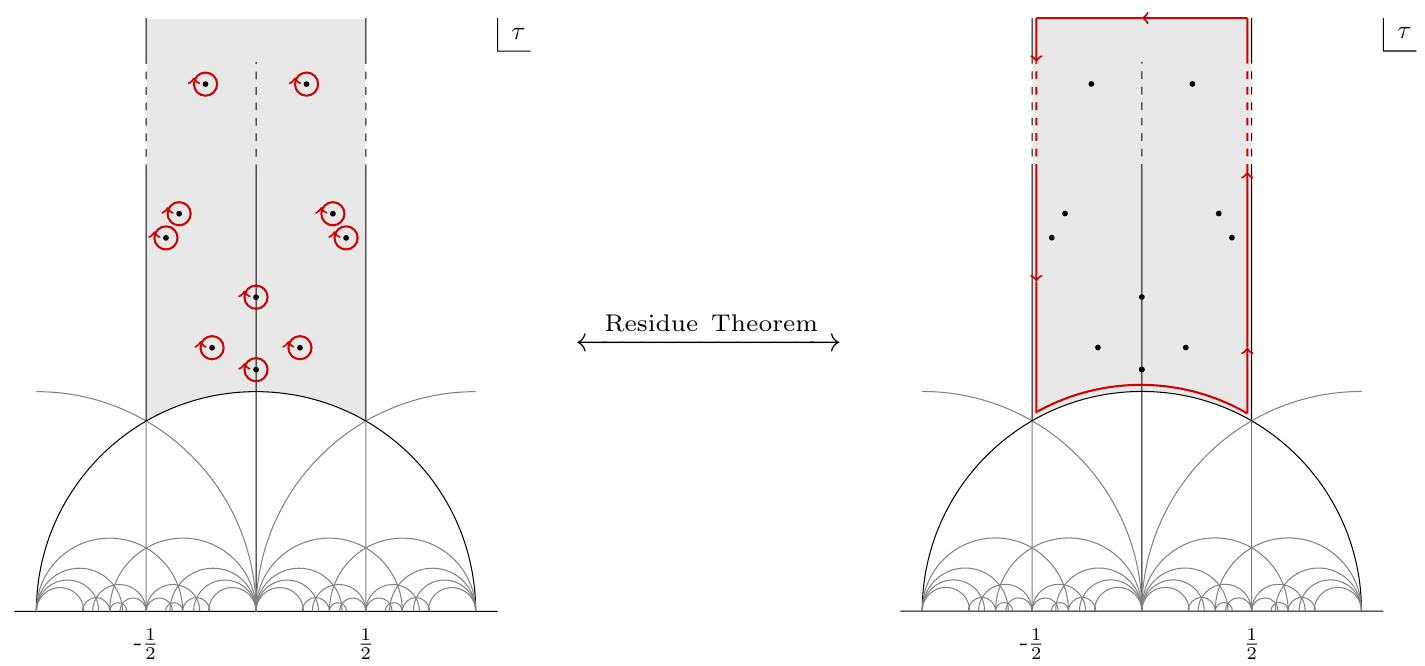}
 \caption{A graphic depiction of the residue theorem on the fundamental domain for $g=1$. Illustrated on the left is the localisation of the integrand on solutions to $u=0$ on the support of the remaining scattering equations. This is \emph{equal} to the integrand, now localised on the boundary $\tau=i\infty$ of the moduli space, with the two expressions related by a residue theorem on the fundamental domain.}
 \label{fig:residue}
\end{center}
\end{figure}

 The key insight  \cite{Geyer:2015bja, Geyer:2015jch} is that the full correlator can,  via residue theorems, be localised on the boundaries of the moduli space, corresponding  geometrically to Riemann spheres with pairs of identified points called a node; these are  \emph{non-separating degenerations}. At one loop, this is achieved by a residue theorem on the moduli space $\mathfrak{M}_{1,n}$, trading the localisation on one of the higher-genus scattering equations, $u=0$, for a localisation on the non-separating boundary divisor $\tau =i\infty$.  This residue theorem, illustrated in \cref{fig:residue},  relies on three properties of the integrand: modular invariance,  complete localisation on the scattering equations, and the presence of a simple pole at the non-separating boundary divisor. Once localised on the nodal sphere, the integrand simplifies drastically, resembling closely a forward limit of an $(n+2)$-particle tree amplitude.

To illustrate this further, consider the integrand $\mathfrak{I}_n^{(1)}$ with the change of variables $q=e^{2i\pi \tau}$, designed to manifest the pole at the non-separating divisor $q=0$. Introducing also a short-hand for better readability, the residue theorem implies that
\begin{equation}
 \mathfrak{I}_n^{(1)}
 := \int_{\mathfrak{M}_{1,n}}\!\!\frac{dq}{q}\, \bar\delta(u)\,\mathcal{I}(q) 
 \stackrel{\mathrm{RT}}{=} \int_{\mathfrak{M}_{1,n}}\!\!\frac{dq}{u}\, \bar\delta(q)\,\mathcal{I}(q)
 = \frac{1}{\ell^2}\int_{\mathfrak{M}_{0,n+2}}\hspace{-15pt}\mathcal{I}(0)\,.
\end{equation}
The first equality is the definition of the short-hand, all remaining terms in \eqref{eq:integrand_g} have been absorbed into $\mathcal{I}(q) $. The residue theorem, in the second equality, relates this expression to the boundary divisor $q=0$, and in the final equality we have used that $u=\ell^2$ on the nodal sphere. Thus, contrary to the string,  the integrand $\mathfrak{I}_n^{(1)}$ is localised on the boundary divisor, reflecting the field-theory status of the ambitwistor string.

\begin{figure}[ht]
\begin{center}
 \includegraphics[width=14cm]{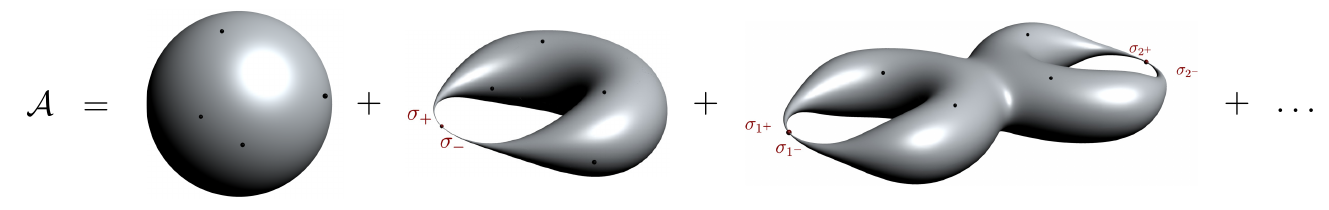}
 \caption{Nodal expansion of the amplitude. The $g$-loop integrand can be expressed as a fully localised integral over  a $g$-nodal sphere.}
 \label{fig:amplitude_nodal}
\end{center}
\end{figure}

While more involved, the argument can be extended to two loops \cite{Geyer:2016wjx, Geyer:2018xwu}, suggesting that the loop expansion in the ambitwistor string has two equivalent representations:  an expansion in the worldsheet genus as in \cref{fig:amplitude_WS}, and a nodal expansion as illustrated in \cref{fig:amplitude_nodal}.

\subsection{The integrand on the nodal sphere}
On the nodal sphere, the one-loop integrand takes the following simple form:
\begin{equation}
{\mathfrak I}^{(1)}_n= \frac{1}{\ell^2}\int_{\mathfrak{M}_{0,n+2}}\hspace{-20pt}\ d\mu_{1,n}^{(\mathrm{nod})}\,\cI^{(1)}\,,
\qquad\qquad d
\mu_{1,n}^{(\mathrm{nod})}\equiv\frac{\prod_{\sA} \bar\delta\big(\mathcal{E}_\sA^{(\mathrm{nod})}\big)\, d\sigma_\sA}{\vol\,\SL(2,\C)\times\C^3}\,.\label{eq:loopint}
\end{equation}
Here the index $A$ runs over the particle labels $i=1,\dots,n$, as well as the two marked points $\sigma_+$ and $\sigma_-$ describing the node. The \emph{nodal scattering equations} $\mathcal{E}_\sA^{(\mathrm{nod})}$, given by the torus scattering equations $\mathcal{E}_i^{(1)}$ on the non-separating boundary,  closely resemble the tree-level scattering equations for $n+2$ particles in a forward limit,
\begin{align}\label{equ:nodalSE}
 &\mathcal{E}_\pm^{(\mathrm{nod})}=\pm \sum_{i} \frac{\ell\cdot k_i}{\sigma_{\pm i}}\,, &&
 \mathcal{E}_i^{(\mathrm{nod})}=\frac{k_i\cdot \ell}{\sigma_{i+}}-\frac{k_i\cdot \ell}{\sigma_{i-}} +\sum_{j\neq i} \frac{k_i\cdot k_j}{\sigma_{ij}}\,.
\end{align}
Geometrically, they impose that the quadratic differential $\mathfrak{P}^{(1)}=P^2(\sigma)-\ell^2\,\omega_{+-}^2$ vanishes globally on the sphere, where $P^\mu$ is now given by
\begin{equation}
P^\mu(\sigma)=\ell^\mu\omega_{+-}(\sigma)+\sum_{i=1}^n \frac{k_i^\mu\, d\sigma}{\sigma-\sigma_i}\,,
\qquad
\omega_{+-}(\sigma) = \frac{\sigma_{+-}\, d\sigma}{(\sigma-\sigma_+)(\sigma-\sigma_-)}\,.
\label{eq:P2_1loop}
\end{equation}
This is a consequence of the residue theorem, which traded precisely the constraint $u=0$ for the localisation on the boundary divisor. As expected, the integrand on the nodal sphere is thus localised on $P^2 = u\,  \omega_{+-}^2 = \ell^2\, \omega_{+-}^2$. The nodal scattering equations can then be written compactly as 
\begin{equation}
 \mathcal{E}_\sA^{(\mathrm{nod})} = \mathrm{Res}_\sA\,\mathfrak{P}^{(1)}= \mathrm{Res}_\sA\big(P^2-\ell^2\,\omega_{+-}^2\big)\,.
\end{equation}

On the nodal sphere, the double copy relations provide an extraordinary `free lunch theorem': although the genus-one representation of the loop integrand only exists for 10d type II supergravity (via the RNS ambitwistor string), nodal sphere expressions exist in any dimension $D\leq 10$, as well as for super Yang-Mills \cite{Geyer:2015bja}. This is achieved by combining the forward limit interpretation of the nodal sphere formul{\ae} with the insights from the \emph{double copy}, which motivates worldsheet  integrands of the form
\begin{align}\label{eq:dc_int_nodal}
 \cI^{(1)}_{\text{sYM}} &=\cI^{(1)}_{\text{col}}\,\cI^{(1)}_{\text{susy}}\,, & \cI^{(1)}_{\text{sugra}} &=\cI^{(1)}_{\text{susy}}\,\widetilde\cI^{(1)}_{\text{susy}}\,.
 \end{align}
 The `kinematic' half-integrand $\cI^{(1)}_{\text{susy}}$ is the nodal sphere limit of the chiral loop integrand $I_n^{\scalebox{0.6}{chi}}$ and can be rigorously derived, whereas  the forward-limit interpretation of the nodal sphere formalism suggests that its colour counterpart can be obtained by `gluing' the colour indices of the loop punctures in the tree-level expression. This leads to a cyclic sum over Parke-Taylor factors with the colour `running in the loop', analogous to the illustration in \cref{fig:loopstotrees}. The half-integrands thus take the form \cite{Geyer:2015jch}
 \begin{align}\label{eq:half_int_nodal}
 \cI^{(1)}_{\text{susy}} &= \cI^{(1)}_{\text{NS}}+ \cI^{(1)}_{\text{R}}\,,
 &
  \cI^{(1)}_{\text{col}} &=\sum_{\rho\in S_n}\frac{
\tr\left(T^{\rho(a_1)}...T^{\rho(a_n)}\right)}{\sigma_{+\,\rho(1)}\sigma_{\rho(1)\,\rho(2)}\dots\sigma_{\rho(n)\,-}\sigma_{-\,+}}\, .
\end{align}
Due to its ambitwistor-string origin, and following the corresponding analysis in the superstring \cite{Tourkine:2012vx}, contributions from individual GSO sectors  can be identified in $\cI^{(1)}_{\text{susy}}$;
\begin{align}
 \cI^{(1)}_{\text{NS}} =\sum_r\pf'\big(M_{\text{NS}}^r\big)\,, \qquad \cI^{(1)}_{\text{R}} =-\frac{c_D}{\sigma_{+-}^2}\ \pf\big(M_2\big)\,.
 \end{align}
 In the Ramond contribution, $c_D$ is a dimension-dependent constant, and the matrix $M_2$    is defined as on the torus \eqref{eq:int_1loop}, but with the nodal sphere Szeg\H{o} kernel 
 $$
 S_2=\sigma_{ij}^{-1}\left(\sqrt{\frac{\sigma_{i+}\sigma_{j-}}{\sigma_{i-}\sigma_{j+}}}+\sqrt{\frac{\sigma_{i-}\sigma_{j+}}{\sigma_{i+}\sigma_{j-}}}\right).
 $$
   The NS-contribution  $\mathcal{I}_{\text{NS}}^{(1)}$ manifests the forward-limit interpretation, 
 \begin{equation}\label{eq:MNS}
 M_{\text{NS}}^r = M_{n+2}^{\text{tree}}\;\Bigg|_{\,{\ell}^2=0\,,\;\epsilon_{+}=\epsilon^r\,,\;\epsilon_{-}=(\epsilon^r)^\dagger}\,,
\end{equation}
where the additional particles at the nodal points $\sigma_{\pm}$ have back-to-back momenta $\pm \ell$, and the sum runs over a basis $\epsilon^r$ of polarisation vectors for these two  particles. 

 Similar representations exist for NS-NS-gravity and pure Yang-Mills theory in various dimensions \cite{Geyer:2015jch}, obtained by replacing $\cI^{(1)}_{\text{susy}}$ with $\cI^{(1)}_{\text{NS}}$. The forward-limit has also been successfully used to construct nodal sphere representation for other theories, including the bi-adjoint scalar \cite{He:2015yua, Cachazo:2015aol, Feng:2016nrf, Feng:2019xiq}  and Einstein-Yang-Mills theory \cite{Porkert:2022efy}, and played an important role in gaining a better understanding of nodal scattering equations \cite{Cachazo:2015aol}  in non-supersymmetric theories.
 
 It is striking that in all of these formul{\ae}, the integrand has the same simplicity as a tree-level amplitude, and is in particular  a rational function of the kinematic data. The residue theorem thus resolves the puzzle of how a field theory integrand can emerge from the complicated higher-genus expressions of the last section. \footnote{At one loop, this is strongly reminiscent of the Feynman Tree Theorem \cite{Feynman:1963ax, Caron-Huot:2010fvq}, but the nodal representations extend to higher loop order, as we will see in \cref{sec:two-loops}.}

 \subsection{Representation of the loop integrand}
The remarkable similarity of the loop integrand $\mathfrak{I}_n^{(1)}$  to  tree-amplitudes in a forward limit has another important consequence: after evaluating the moduli space integrals, the loop integrand appears in a non-standard representation on momentum space. This is already evident from the general form of the \eqref{eq:loopint}, which involves inverse quadratic powers of the loop momentum only via the overall factor $\ell^{-2}$, whereas the nodal scattering equations \eqref{equ:nodalSE} depend linearly on $\ell$. This intuition can be made precise by factorisation arguments analogous to \S\ref{sec:proof}, showing that the integrand contains poles at $2\ell\cdot K + K^2$ instead of the conventional Feynman loop propagators $(\ell+K)^2$, \cite{Geyer:2015jch}.

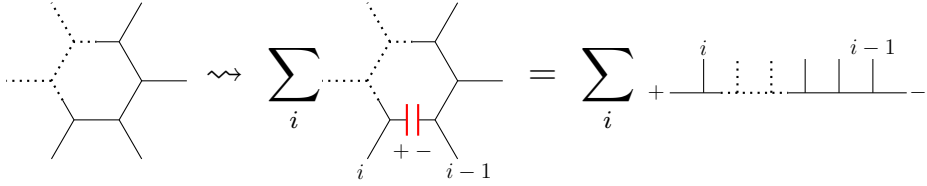
\begin{figure}[ht]\vspace{7pt}
\begin{center}
 \begin{tikzpicture}[scale=0.3]
 \draw (-0.5,0.866) -- (0,0) -- (1,0) -- (2,0) -- (3,1.732) -- (2,3.464) -- (1,3.464);
 \draw[dotted, thick] (-0.5,0.866) -- (-1,1.732) -- (0,3.464) -- (1,3.464);
 \draw (0,0) -- (-1,-1.732);
 \draw[dotted, thick] (-1,1.732) -- (-3,1.732);
 \draw[dotted, thick] (0,3.464) -- (-1, 5.196);
 \draw (3,1.732) -- (5,1.732);
 \draw (2,0) -- (3,-1.732);
 \draw (2,3.464) -- (3, 5.196);
 
 \node at (8.4,1.4) {\scalebox{1.2}{$\leadsto\,\,\displaystyle\sum_i$}};
 \draw (13.5,0.866) -- (14,0) -- (14.75,0);
 \draw (15.25,0) -- (16,0) -- (17,1.732) -- (16,3.464) -- (15,3.464);
 \draw[dotted, thick] (13.5,0.866) -- (13,1.732) -- (14,3.464) -- (15,3.464);
 \draw (14,0) -- (13,-1.732);
 \draw[dotted, thick] (13,1.732) -- (11,1.732);
 \draw[dotted, thick] (14,3.464) -- (13, 5.196);
 \draw (17,1.732) -- (19,1.732);
 \draw (16,0) -- (17,-1.732);
 \draw (16,3.464) -- (17, 5.196);
 \draw[red,thick] (14.75,-0.7) -- (14.75,0.7);
 \draw[red,thick] (15.25,-0.7) -- (15.25,0.7);
 \node at (14.5, -1.4) {\scalebox{0.7}{$+$}};
 \node at (15.5,-1.4) {\scalebox{0.7}{$-$}};
 \node at (12.7,-2.3) {\scalebox{0.7}{$i$}};
 \node at (17.5,-2.3) {\scalebox{0.7}{$i-1$}};
 
 \node at (22.5,1.4) {\scalebox{1.2}{$=\,\,\displaystyle\sum_i$}};
 \draw (26.4,1.2) -- (28.65,1.2);
 \draw (27.9,1.2) -- (27.9,2.7);
 \draw[dotted,thick] (28.65,1.2) -- (31.65,1.2);
 \draw (31.65,1.2) -- (36.9,1.2);
 \draw[dotted,thick] (29.4,1.2) -- (29.4,2.7);
 \draw[dotted,thick] (30.9,1.2) -- (30.9,2.7);
 \draw (32.4,1.2) -- (32.4,2.7);
 \draw (33.9,1.2) -- (33.9,2.7);
 \draw (35.4,1.2) -- (35.4,2.7);
 \node at (25.8, 1.2) {\scalebox{0.7}{$+$}};
 \node at (37.4,1.2) {\scalebox{0.7}{$-$}};
 \node at (27.9,3.2) {\scalebox{0.7}{$i$}};
 \node at (35.4,3.2) {\scalebox{0.7}{$i-1$}};
\end{tikzpicture}
\end{center}\vspace{-20pt}
\caption{Interpretation of the $\mathfrak{I}_{\mathrm
{lin}}$ representation of loop integrands as $(n+2)$-particle tree diagrams, summed over different ways of `cutting open' the loop.}
\label{fig:loopstotrees}
\end{figure}

Fortunately this novel,  `linear' representation can be obtained from the standard loop integrand  by a simple prescription, based on a deformation and residue theorem reminiscent of the BCFW recursion, \cite{Baadsgaard:2015twa}.  This can be achieved as follows: shift the loop momentum in the standard representation $\mathfrak{I}_{\scalebox{0.7}{std}}$ by $\ell\rightarrow \tilde\ell=\ell+\eta$, where $\eta$ points in some auxiliary dimension such that  $\ell\cdot\eta=k_i\cdot\eta=\epsilon_i\cdot\eta=0$, and the Lorentz invariants are unaffected except for $\ell^2\rightarrow \ell^2+\eta^2\equiv\ell^2+\zeta$. Cauchy's residue theorem then relates the $\mathfrak{I}_{\scalebox{0.7}{std}}$, expressed now as the residue at $\zeta=0$, to a sum of terms where all but one of the propagators are linear. A further shift in the loop momentum $\ell\rightarrow \ell-K_a$, where $K_a=\sum_{i\in I_a}k_i$ is the sum of external momenta in a propagator $D_a$, brings all remaining quadratic propagators into the form $\ell^{-2}$, and gives the linear representation of the loop integrand;
\begin{equation}\label{eq:IlinQ}
 \mathfrak{I}_{\scalebox{0.7}{std}}=\sum_\Gamma\frac{N\big(\ell,\ell^2\big)}{\prod_{a\in\Gamma}D_a}
 \quad \leadsto \quad
\mathfrak{I}_{\scalebox{0.7}{lin}}=\frac{1}{\ell^2}\sum_\Gamma\sum_{a\in\Gamma}\frac{N\big(\ell-K_a,\,-2\ell\cdot K_a +K_a^2\big)}{\prod_{b\neq a}(D_b-D_a)\Big|_{\ell\rightarrow \ell-K_a}}  \,.
\end{equation}
Here $D_a=(\ell+K_a)^2$ are the standard Feynman loop propagators, and we have kept the  dependence on $\ell^2$ explicit  in the numerators $N(\ell,\ell^2)$  for better readability. At this stage, it is easily verified that the remaining propagators in $\mathfrak{I}_{\scalebox{0.7}{lin}}$ are of the form $2\ell\cdot K+K^2$ expected from the nodal scattering equations. 
\footnote{In the simplest case when the numerators are independent of $\ell^2$, the residue theorem reduces to repeated partial fraction identities, followed by  shifts in the loop momentum $\ell\rightarrow \ell-K_a$ as above;
\begin{equation}
 \frac{1}{\prod_a D_a}=\sum_a \frac{1}{D_a\prod_{b\neq a}(D_b-D_a)}\,,\qquad \text{where } D_a=(\ell+K_a)^2\text{ and }K_a=\sum_{i\in I_a}k_i\,.
\label{eq:partialfrac}
\end{equation} }
Schematically, the sum over different propagators $a$ in $\mathfrak{I}_{\scalebox{0.7}{lin}}$ can be given an interpretation as different ways of `cutting open' the loop, with each term associated to a tree-diagram involving two back-to-back on-shell momenta $\pm\tilde\ell =\pm( \ell+\eta)$; see \cref{fig:loopstotrees}.

The procedure \eqref{eq:IlinQ} serves as an algorithm for deriving the linear representation $\mathfrak{I}_{\scalebox{0.7}{lin}}$ from a standard integrand $\mathfrak{I}_{\scalebox{0.7}{std}}$. To date, there exists no general algorithm for the reverse direction, impacting our ability to apply established integration techniques as in review chapters 1 and 3 \cite{chapter1, chapter3} for its evaluation. On the other hand, its non-standard structure also has clear advantages: as we shall see below, the forward-limit structure facilitates the extension of tree-level results to loop level, impacting the double copy at loop level (\S\ref{sec:DC_loops}), and allowing for scattering equations-based formul{\ae} for the bi-adjoint scalar at one loop  \cite{He:2015yua, Cachazo:2015aol, Feng:2016nrf, Feng:2019xiq}. Following an alternative direction, there has also been work on obtaining standard loop integrands with Feynman propagators from the scattering equations formalism in  \cite{Gomez:2017lhy, Gomez:2017cpe, Ahmadiniaz:2018nvr} and  \cite{Farrow:2020voh}, but at the cost of more complicated expressions, and the origin of these formul{\ae} from the ambitwistor string correlator remains unclear.

\subsection{Double copy at loop level}
\label{sec:DC_loops}

At loop level, the double copy is conjectural, but explicit constructions exist for loop intgrands in a variety of theories, \cite{Bern:2010ue,Carrasco:2011mn,Bern:2012uf, Bern:2012cd, Bern:2013qca, Bern:2013uka, Bern:2014sna, Mafra:2015mja, Johansson:2017bfl}, see also the review chapter 2 \cite{Chapter2}. In the ambitwistor string, the first incarnation of the double copy at loop level can already be found in the structure of the worldsheet integrands \eqref{eq:dc_int_nodal}. Extending the corresponding tree-level result, the BCJ relations also embed straightforwardly  \cite{He:2016mzd}, 
\begin{equation}
 \sum_{j=1}^{n-1}\frac{\ell\cdot k_{12...j}}{\sigma_{12}...\sigma_{j+}\sigma_{+\,(j+1)}...\sigma_{n-}\sigma_{-1}}=0 \,,\qquad\text{mod }E_a^{(\text{nod})}\,.
\end{equation}
We can make contact with the double copy on momentum space by expanding
both the Yang-Mills integrand $\mathfrak{I}_{\text{YM}}$ and the gravity integrand $\mathfrak{I}_{\text{grav}}$ in a  Dixon-Del Duca-Maltoni (DDM) half-ladder basis, 
\begin{align}\label{equ:expansionloop}
 \mathfrak{I}_{\text{YM}}&=\sum_{\rho\in S_n} c(+,\rho,-)\, \mathfrak{I}_{\text{YM}}(+,\rho,-) \,,
&
 \mathfrak{I}_{\text{grav}}&=\sum_{\rho\in S_n} N(+,\rho,-)\, \mathfrak{I}_{\text{YM}}(+\rho,-) \,,
\end{align}
where $\mathfrak{I}_{\text{YM}}(+,\rho,-)$ are colour-ordered Yang-Mills integrands. Both the colour factors
\begin{equation}
c(+,\rho,-) = f^{a_+a_{\rho(1)}b_1} \, f^{b_1a_{\rho(2)}b_2} \cdots f^{b_{n-1}a_{\rho(n)}a_-}\, \delta^{a_+a_-} \,,
\end{equation}
and the kinematic numerators $N(+,\rho,-)$ are associated 
to cubic diagrams forming a `half-ladder', with legs $+$ and $-$ at opposite endpoints, as on the right of \cref{fig:loopstotrees}.  If such an expansion can be found,
the integrands $\mathfrak{I}_{\text{grav}}$ and $\mathfrak{I}_{\text{YM}}$ are related by replacing the colour factors by the numerators $N$, and so they satisfy the double copy structure. Such numerators $N(+,\rho,-)$ are also known as BCJ numerators or master numerators, since they generate the  numerators for all other diagrams by Jacobi relations. For the ambitwistor string integrands $\mathcal{I}^{(1)}$, this expansion becomes
\begin{align}
& \cI_{\text{col}}^{(1)} \; = \sum_{\rho\in S_{n}} 
\frac{ c(+,\rho,-) 
}{\sigma_{+\rho(1)}...\sigma_{\rho(n)-}\sigma_{-+}} \,,  
&
& \cI_{\text{kin}}^{(1)} \;  \cong \sum_{\rho\in S_{n}} 
\frac{ N(+,\rho,-)
}{\sigma_{+\rho(1)}...\sigma_{\rho(n)-}\sigma_{-+}} \,,
\label{equ:expansionpfaffianloop}
\end{align}
and is guaranteed to exist due the one-loop KLT orthogonality \cite{He:2016mzd, He:2017spx} 
on the support of the  nodal scattering equations,  as expressed by the $\cong$-symbol. Various strategies have been successfully used to determine the BCJ numerators $N$ for supersymmetric and non-supersymmetric theories  
 \cite{He:2016mzd, He:2017spx, Geyer:2017ela, Edison:2020uzf}, often generalising tree-level methods by exploiting the forward-limit structure of the loop integrand, and some constructions have been extended to two loops \cite{Geyer:2019hnn}. The relative ease with which   BCJ numerators can be constructed in the linear, ambitwistor-string inspired integrand representation stands in stark contrast to the status in the standard representation, where  serious obstacles arise already for six external particles at one loop \cite{Mafra:2014gja, Berg:2016fui}, see also  \cite{He:2016mzd, He:2017spx} for a concise juxtaposition.

\subsection{Two loops}
\label{sec:two-loops}
The nodal sphere formalism has been successfully extended to two loops \cite{Geyer:2016wjx, Geyer:2018xwu}. In this case, the residue theorem is more subtle, but it remains true that the full integrand localises on the maximal non-separating boundary of the moduli space, corresponding to a bi-nodal sphere,  parametrised by four `loop marked points', one pair per node,
 \begin{align}
\label{eq:assamp}
 \mathfrak{I}^{(2)}_4 & = \frac{\mathcal{K}\tilde{\mathcal{K}}}{\prod_I(\ell^I)^2}  \int_{{\mathfrak M}_{0,4+2g}} \hspace{-10pt} c^{(g)}\big({\mathcal J}^{(g)}\mathcal{Y}^{(g)}\big)^2  \prod_{\sA=1}^{4+4}{}'\bar\delta(\mathcal{E}_\sA)\   \,.
\end{align}
The resulting integrand formula, presented here for $n=4$ particles for simplicity,  takes the form of an integral over the  moduli space ${\mathfrak M}_{0,n+2g}$, fully localised on solutions to the nodal scattering equations $\mathcal{E}_\sA=0$. While the structure is reminiscent of the one-loop case, new features appear as well; in particular the  factors $c^{(g)}$ and ${\mathcal J}^{(g)}$ arise from the degeneration of ${\mathfrak M}_{g,n}$ to ${\mathfrak M}_{0,n+2g}$. We briefly discuss these ingredients below, more detailed expositions, as well as $n$-point formul{\ae} for supergravity and super Yang-Mills theory (constructed again using the double copy) can be found in the original paper \cite{Geyer:2018xwu}.

\begin{itemize}
\item \emph{Moduli:} The residue theorem localises the integrand on the non-separating boundary, where $q_{II}=e^{i\pi\Omega_{II}} =0$. In this limit, the remaining moduli, given by the off-diagonal components of the period matrix,  become cross-ratios of the nodal marked points $\sigma_{I^\pm}$, 
 \begin{equation}
 \label{eq:qij}
q_{IJ}=e^{2i\pi\Omega_{IJ}}=\frac{\sigma_{I^+J^+}\sigma_{I^-J^-}}{\sigma_{I^+J^-}\sigma_{I^-J^+}} \,.
\end{equation}
The measure is then naturally expressed in terms of all marked points (including the nodes) modulo M\"{o}bius transformation, leading to a  Jacobian  ${\mathcal J}^{(g)}$ with
\begin{equation}
\label{eq:defJ}
 \prod_{I<J}\frac{dq_{IJ}}{q_{IJ}} 
 = \frac{{\mathcal J}^{(g)}}{\mathrm{vol\; SL}(2,\mathbb{C})}
 \,, \quad {\mathcal J}^{(g)}=\frac{1}{\sigma_{1^+2^+}\sigma_{1^+2^-}\sigma_{1^-2^+}\sigma_{1^-2^-}}\prod_{I^\pm}d\sigma_{I^\pm}\,.
\end{equation}
The analogous change of variables in the scattering equations  results in another copy of the same Jacobian factor.

The cross-ratio factor  $\,c^{(2)} = 1/(1-q_{12})$ originates in the degeneration of the moduli space ${\mathfrak M}_{2,n}$ to ${\mathfrak M}_{0,n+4}$; to be precise, from mapping the last modular parameter $q_{12}$ to the nodal sphere \cite{Geyer:2018xwu}. It can be given a concrete physical interpretation in projecting out unphysical poles from the integrand $\mathfrak{I}_n^{(2)}$.
 \item \emph{Nodal scattering equations:} On the nodal sphere, $P_\mu$ takes the following form;
 \begin{equation}
 P_\mu(\sigma) = \ell^I_\mu \,\omega_{I^+I^-}(\sigma) + \sum_i \frac{k_{i\mu}}{\sigma-\sigma_i} \,d\sigma \,, 
\end{equation}
where $\omega_{I^+I^-}(\sigma) $ are the genus-two  holomorphic Abelian differentials in the maximal non-separating degeneration. In this limit, these differentials  acquire simple poles at the corresponding nodes: 
 \begin{equation}\label{eq:abelian-diff_g2}
\omega_I=\frac{\omega_{I^+I^-}}{2\pi i}\,, \quad 
\omega_{I^+I^-}(\sigma)= \frac{(\sigma_{I^+}-\sigma_{I^-})\, d\sigma}{(\sigma-\sigma_{I^+})(\sigma-\sigma_{I^-})}\,,
\end{equation}
The nodal scattering equations can then be compactly expressed as the  vanishing of a meromorphic quadratic differential $\mathfrak{P}^{(g)}$ with only simple poles, 
\begin{equation}
 \mathcal{E}_\sA =\mathrm{Res}_{\sigma_{\!A}}\mathfrak{P}^{(g)} \,,\qquad
 \mathfrak{P}^{(2)}=P^2 - (\ell^I \!\omega_{I^+I^-})^2 +(\ell_1^2+\ell_2^2)\,\omega_{1^+1^-}\omega_{2^+2^-}\,.
\end{equation}
Note in particular the last term, a novel feature at genus two that plays a crucial role in obtaining the correct loop propagators. \footnote{and is closely related to the absence of a straightforward Feynman tree-theorem at two loops.} 
\item \emph{Integrand:} The chiral integrand is defined straightforwardly by the nodal sphere-limit of the genus-two expression, $\mathcal{Y} = \mathcal{Y}^{(2)}\big|_{\mathrm{nodal}}$, and can be calculated using \eqref{eq:abelian-diff_g2}.
\end{itemize}

\subsection{Further topics}
The nodal sphere formulation of loop integrands proved to be a starting point for many further exciting avenues of research. 

\begin{itemize}
 \item At the level of the worldsheet model, the simple structure of one-loop correlators, supported on a nodal sphere is reflected by the presence of a so-called `gluing operator' in the ambitwistor string \cite{Roehrig:2017gbt}. 
 This gluing operator $\Delta$ encodes the propagator of the target-space field theory, and is thus a BRST-invariant but non-local worldsheet operator. Genus zero correlators with an insertion of $\Delta$ directly give the one-loop integrand formul{\ae} \eqref{eq:loopint}  localized on the nodal sphere, without need for further simplifications. Extensions to higher loops are currently not known.
 \item Focusing on the integrand expressions, there has been tremendous progress on extending many of the tree-level evaluation techniques  to loop level,  \cite{Baadsgaard:2015hia, Zlotnikov:2016wtk,Gomez:2016cqb, Chen:2016fgi}.
 \item The nodal sphere, and in particular the forward limit structure of the integrand, also inspired loop formul{\ae}  in the twistorial models \cite{Wen:2020qrj}. These were obtained from a forward limit of the 6d spinorial tree-level amplitudes mentioned in \cref{sec:twistor-models}.
 \item Very recently, a proposal has also appeared for one-loop correlators in massive $\phi^4$ theory on de Sitter spacetime, based on the nodal sphere  \cite{Gomez:2021ujt}. This builds on earlier work expressing (tree-level) de Sitter `cosmological correlators'  as worldsheet integrals, supported on so-called cosmological scattering equations, which are now differential operators expressed as functions of the conformal generators  \cite{Gomez:2021qfd}.
 \item Finally, the ambitwistor string progress at loop level has also inspired calculations and proposals in the full superstring. In ref.~\cite{Edison:2021ebi}, the authors used an ambitwistor-string-inspired method, based on forward limits of the moduli space integrals, to construct one-loop matrix elements with insertions of operators $D^{2k}  F^{n}$ and $D^{2k} R^n$ in the tree-level effective action. Progress has also been made at higher loop orders, where the double copy and the close relation between the ambitwistor string and the superstring chiral integrands has been used to propose a formula for the three-loop four-particle superstring integrand \cite{Geyer:2021oox}. This is achieved by constructing first an expression on the three-nodal sphere using BCJ numerators, and then lifting this to a fully modular invariant proposal for the superstring chiral integrand on a $g=3$ Riemann surface. 
\end{itemize}


\section{Frontiers}\label{Frontiers}

As we have seen, ambitwistor-strings give one of the most direct routes from a physical theory to compact formul{\ae} for tree amplitudes and loop integrands.  Despite these successes, many open questions remain and it is questionable as to whether  it will one day be possible to understand these worldsheet models as providing  secure  basic formulations of physical theories.  To consolidate them, more work needs to be done to relate them to more standard fully nonlinear formulations of physical theories, either via field theory, string theory or   holography. We briefly expand on these connections.

\subsection{Curved backgrounds}
Amplitudes on curved backgrounds are a relatively new subject with the frontier being, until recently, at three points at tree level.  They provide  a stepping stone to connect with conventional nonlinear field theory. 
Spaces of complexified null geodesics make good sense on an analytic curved space with metric $g(x)_{\mu\nu}$ \cite{LeBrun:1983} and one can ask whether ambitwistor  strings can be defined on such curved ambitwistor spaces.  In particular in the RNS models of \S\ref{sec:models} we can replace $P^2\rightarrow H=g^{\mu\nu}(x)P_\mu P_\nu+\ldots $.  It was shown in \cite{Adamo:2014wea} that, in the type II case,  curved analogues of the constraints $H$, $G=\Psi\cdot P$ and $\tilde G=\tilde \Psi\cdot P$ can be constructed so that they satisfy the flat space OPEs iff they are obtained from a solution to the NS-NS-sector of 10d type II supergravity. This can be used to construct vertex operators and amplitudes at three points on a plane wave \cite{Adamo:2017sze, Adamo:2018ege}, and one can similarly encode  the Yang-Mills equations in the heterotic model \cite{Adamo:2018hzd}. This provides a completely different perspective to that pursued in the 1970's and 1980's when space-time field equations in 4d were shown \cite{Isenberg:1978kk, Baston:1987av, LeBrun:1991jh} to correspond to the existence of formal neighbourhoods of $\A$ inside $\P\T\times \P\T^*$ or supersymmetric extensions \cite{ Witten:1978xx,Witten:1985nt}.  In the ambitwistor string, the field equations are encoded in the quantum consistency of the worldsheet model on a curved ambitwistor space.   In all these cases and unlike the conventional string, one can use field redefinitions to make the gauge-fixed action linear, so that correlators are relatively easy to compute.  However, the complexity of the curved background constraints makes it problematic to identify the generic integrated vertex operators.

The 4d twistor models and twistor-strings can also be defined on curved backgrounds built from curved twistor spaces. These are equivalent to curved but self-dual Yang-Mills or Einstein backgrounds, which are integrable. Exploiting this integrability,  one can construct amplitudes of arbitrary multiplicity, \cite{Adamo:2021bej, Adamo:2020yzi,Adamo:2022mev} with explicit formul{\ae} for certain classes of backgrounds.

\subsection{Relationship with conventional, null and chiral strings}\label{chiral}
All the amplitude formul{\ae} we have discussed localize on the solutions to the scattering equations.
These equations were first obtained by Fairlie \cite{ Fairlie:1972zz, Fairlie:2008dg} in a semiclassical study of string solutions, and made famous in the work of  Gross and Mende~\cite{Gross:1987ar,Gross:1987kza}, where they were shown to govern the string path integral in the limit of high energy scattering at fixed angle ($s\gg1/\alpha'$).  The scattering equations thus play a prominent role both in the $\alpha'\rightarrow \infty$ limit and in the field theory limit $\alpha'\rightarrow 0$ , albeit indirectly via the ambitwistor string. This clash, with the scattering equations appearing in both the low tension and high tension field theory limit, has so far impeded attempts to connect the ambitwistor string directly with conventional strings. In this context, it is interesting to note that the $\alpha'\rightarrow \infty $ limit can also be understood as  a null limit in which the worldsheet becomes null ruled by null geodesics  \cite{Casali:2016atr}. 


More generally Siegel proposed that so-called chiral string theories could be obtained by flipping certain worldsheet boundary conditions \cite{Siegel:2015axg}, so that both left-moving and right-moving modes in the conventional string become holomorphic on the worldsheet in the sectorized  chiral string.   
A bosonic such model had already been introduced in \cite{Hohm:2013jaa}, and pure spinor versions discussed in \cite{Azevedo:2017yjy}.  In these models both the left moving and right-moving Virasoro constraints become holomorphic but commute with each other and each satisfies  the holomorphic Virasoro algebra.  Ambitwistor have a similar character in the sense that $X^\mu$ and $\int^\sigma P^\mu$ can be thought of as being independent holomorphic functions on the worldsheet playing roles as different combinations of left and right movers. However, the ambitwistor-string only contains one holomorphic Virasoro generator,  $P\cdot \p X$.   The gauged constraint $P^2$  is analogous to difference between the two Virasoro generators, but has trivial OPE with itself, even on a curved background. This reflects the degeneration of the two copies of the Virasoro algebra to a Galilean Conformal Algebra \cite{Casali:2016atr, Casali:2017zkz}. On the other hand,  in the model of \cite{Hohm:2013jaa},   the gauged constraint $P^2$ is replaced by a more general quadratic expression in $P$ and $\p X$ 
\begin{equation}
P^2\rightarrow \cH:=A_{\mu\nu}(X)P^\mu P^\nu + B_{\mu\nu}P^\mu\p X^\nu+C(x)_{\mu\nu}\p X^\mu\p X^\nu\, ,  
\end{equation}
that is constrained to obey a nontrivial OPE.  The imposition of these OPEs yields field equations for the background with a finite number of $\alpha'$ corrections.   However, since the OPE of $\cH$ with itself is nontrivial, there  no longer appears to be a reduction to ambitwistor space nor localization on the scattering equations.  
  The connections between sectorized strings, null strings and ambitwistor strings are now well studied in different models \cite{Casali:2016atr, Casali:2017zkz, Azevedo:2017yjy, Azevedo:2019zbn}, see also \cite{Jusinskas:2021bdj, Lize:2021una}. For example, versions of T-duality become possible in sectorized strings whereas they are not in the ambitwistor-string \cite{Casali:2017mss}.

The amplitude formul{\ae} to which such sectorized chiral strings  give rise  appear to be problematic.  The Koba-Nielsen factor of the conventional string, consisting of a product $\prod_{i<j}|\sigma_{ij}|^{\alpha' s_{ij}}$, is replaced by  a product $\prod_{i<j}(\sigma_{ij}/\bar\sigma_{ij})^{\alpha' s_{ij}}$ so that the branching would seem to make the contour prescription  problematic.  This is resolved in the work of Mizera who uses twisted cohomology and residues to define the amplitudes of these theories \cite{Mizera:2019gea}. As $\alpha'\rightarrow \infty$ he shows localization on the scattering equations.

\subsection{Celestial holography and soft theorems}
Celestial holography \cite{chapter11} seeks to understand the S-matrices of massless theories by formulating these theories on the conformal boundary of asymptotically flat space-times,  $\scri$, the  \emph{light-cone at infinity}. This approach emerged from the study of connections between soft theorems for amplitudes and asymptotic BMS symmetries of space-time \cite{Strominger:2013jfa,Strominger:2017zoo}, and aims to establish a holographic dictionary for the S-matrix  from $\scri$ or the celestial sphere  \cite{Pasterski:2017kqt,Pasterski:2021rjz}.  In an asymptotically simple space-time,  all light rays reach $\scri$ and we can represent ambitwistor space $\A=T^*\scri_\C$ as  the cotangent bundle of the complexification of null infinity. Both the RNS and 4d twistor ambitwistor models can be expressed in this representation \cite{Adamo:2014yya, Geyer:2014lca} and from this perspective, the connection between soft theorems and asymptotic symmetries can be understood directly at the level of the vertex operators as generators of supertranslations and super-rotations in their soft limits. The ambitwistor strings then provide an underpinning theory for the generation of the S-matrix now seen as being based at $\scri$.  
The recently discovered gravitational `$w_{1+\infty}$-symmetry' \cite{Strominger:2021lvk} can be seen in this formulation as arising directly from the geometry  of the asymptotic twistor space at $\scri$  \cite{Adamo:2021lrv}.

\section*{Acknowledgements}
This work  was supported  by the European Union's Horizon 2020 research and innovation programme under the Marie Sk\l{}odowska-Curie grant agreement No.~764850 {\it ``\href{https://sagex.org}{SAGEX}''}. 
YG is grateful for the support from the CUniverse research promotion project ``Toward World-class Fundamental Physics'' of Chulalongkorn University (grant reference CUAASC). The research of LJM  is supported by the STFC grant ST/T000864/1.


\appendix

\section{Penrose transform}
The Penrose transform for a general ambitwistor space $\P\A$ maps spin $s$ fields on space-time to elements of $H^1(\P\A,\cO(s-1))$, with the classes corresponding to spin $s$ plane-waves of the form $\epsilon_{\mu_1} \ldots \epsilon_{\mu_s}\e^{ik\cdot X}$ mapping to $
(\epsilon\cdot P)^s \,\bar \delta (k\cdot P) \,\e^{ik\cdot x}$.

Briefly, this follows by taking a cohomology class $\phi\in H^1(\P\A, \cO(s-1))$ on $\P\A$, pulling it back to $p^*\phi$ on $PT^*M|_{P^2=0}$, the lightcone  inside the projective cotangent bundle,  where it becomes necessarily trivial as there is no first cohomology on this space for the given homogeneity weights. 
Thus we can find $g $ of weight $s-1$ on $T^*M|_{P^2}=0$ such that
\begin{equation}\label{ptrans}
p^*\phi=\dbar g\, , \qquad \mbox{e.g.\  for $\phi$ in \eqref{drep} } \qquad g= (\epsilon\cdot P)^s \frac{\e^{ik\cdot X}}{k\cdot P}\, .
\end{equation}
Then we can obtain the field via
\begin{equation}
(\epsilon\cdot P)^s \,\e^{ik\cdot x}= P\cdot \p_X g\, .
\end{equation}
Since  $g$ is defined up to the gauge freedom, $\delta g= (\epsilon\cdot P)^{s-1}\e^{ik\cdot X}$ (or indeed some a general global holomorphic function of $P$),  we  have the usual gauge freedom 
\begin{equation}
\delta (\epsilon_{\mu_1} \ldots \epsilon_{\mu_s} \e^{ik\cdot X})=k_{(\mu_1} \epsilon_{\mu_2}\ldots \epsilon_{\mu_s)}\, \e^{ik\cdot X}.
\end{equation}
This can be proven more abstractly as follows.  The Penrose transform is  the connecting  map 
\begin{equation}\label{conn-hom}
H^0(PT^*M_{P^2=0},\cO(s))/P\cdot \p_X \left(H^0(PT^*M_{P^2=0},\cO(s-1))\right) \stackrel{\delta}\longrightarrow H^1(\P\A,\cO(s-1)),,
\end{equation}
from the long exact sequence in cohomology arising from the short exact sequence:
\begin{equation}\label{SES}
0\rightarrow  \cO(s-1)_{\P\A}\rightarrow \cO(s-1)_{T^*M_{P^2=0}}\xrightarrow{P\cdot \p_X} \cO(s)_{T^*M_{P^2=0}} \rightarrow 0\,, .
\end{equation} 
that defines the functions on $\P\A$, see \cite{ Baston:1987av,Mason:2013sva} for a full discussion.

\section{Conformal Field Theory basics}\label{sec:CFT}
In this appendix, we present a lightning review  of some aspects of chiral two-dimensional conformal field theories (CFTs) particularly relevant for the ambitwistor string. For a more extensive introduction, we refer the interested reader to the multitude of excellent textbooks, for example \cite{Green:1987a,Green:1987b,Polchinski:1998rq,Ketov:1995,DiFrancesco:1997nk,Polchinski:1998rr}.

All ambitwistor string models are two-dimensional CFTs with a local action in terms of a set of fields defined over a closed Riemann surface $\Sigma$, referred to as the worldsheet. These fields can be characterized by their statistics (bosonic vs fermionic) and their conformal weight, as well as additional quantum numbers such as the spin structure at higher genus. The conformal weight of a field $\Phi$ is a pair of half-integers $(h,\bar h)\in\mathbb{Z}/2\mathbb{Z}\times\mathbb{Z}/2\mathbb{Z}$ labeling the transformation properties of $\Phi$ under two-dimensional conformal transformations. \footnote{A field is called \emph{primary} if its conformal weights are well-defined.} 
Since 2d local conformal transformations are equivalent to holomorphic coordinate transformations, we can identify the conformal weight with the form degree. This means that  $\Phi$ is section of $(h,\bar h)$ powers fo the holomorphic and antiholomorphic canonical bundles $K_\Sigma^h\otimes \overline{K}_\Sigma^{\bar h}$, i.e.  
\begin{equation}
 \Phi\in \Omega^0(\Sigma, K_\Sigma^h\otimes \overline{K}_\Sigma^{\bar h})\,.
\end{equation}
Here, negative weights are to be interpreted as sections of the respective tangent bundles, using the isomorphism $K_\Sigma^{-1}\cong T_\Sigma$. Equivalently, $\Phi$ may be expressed locally in affine worldsheet coordinates $z$ as
\begin{equation}
 \Phi(z,\bar z) = \phi_{h,\bar h} \,(dz)^h\, (d\bar z) ^{\bar h}\,.
\end{equation}

In the ambitwistor string, we will only encounter \emph{chiral} CFTs, known as $\beta\gamma$-systems (for bosons) or $bc$-systems (for fermions). We will describe these jointly below, keeping track of the statistics via a variable $\epsilon$, with $\epsilon=-1$ for fermionic statistics, and $\epsilon=1$ for bosons. A chiral CFT is then defined by the action
\begin{equation}
 S=\frac{1}{2\pi}\int b\bar\partial c\,,
\end{equation}
in conformal gauge. The fields $b$ has conformal weight $(h,0)$, giving conformal weight  $(1-h,0)$ to the conjugate field $c$. Fields such as these with conformal weight $\bar h=0$ are often referred to as `left-moving'.  The OPE between these conjugate fields  is 
\begin{equation}
 c(z)\,b(w)\sim\frac{1}{z-w}\,,
 \qquad\qquad
 b(z)\,c(w)\sim-\frac{\epsilon}{z-w}\,.
\end{equation}
A standard calculation gives the holomorphic stress-energy tensor
\begin{equation}
 T_{bc} = -h \, b\partial c+(1-h)\,(\partial b) c\,.
\end{equation}
From this expression, we find the central charge anomaly  as (twice) the coefficient of the fourth order pole in the $T(z) T(w)$ OPE, 
\begin{equation}
 \mathfrak{c}= 2\epsilon \left(6h^2-6h+1\right)\,.
\end{equation}
Finally we note the following useful formula for the number of zero modes $n_b$ and $n_c$ of the two fields on a Riemann surface of genus $g$, derived via the Riemann-Roch theorem,
\begin{equation}
 n_c-n_b = \frac{1}{2}\left(2h-1\right)\chi\,,
 \qquad \mathrm{with}\;\;\chi = 2(1-g)\,.
\end{equation}

\section*{Bibliography}

\bibliography{twistor-bib}
\bibliographystyle{JHEP}

\newcommand{\eprint}[2][]{\href{https://arxiv.org/abs/#2}{\tt{#2}}}

\end{document}